\documentclass[twocolumn,astrosymb,times,twocolappendix]{aastex631}

\newcommand{\mean}[1]{\langle#1\rangle}

\submitjournal{ApJ}

\shorttitle{Deep learning for Protoclusters}
\shortauthors{Takeda et al.}

\usepackage{amsmath}
\usepackage[T1]{fontenc}
\usepackage{bm}

\begin{document}

\title{Mining for Protoclusters at $z\sim4$ from Photometric Datasets with Deep Learning}

\correspondingauthor{Yoshihiro Takeda}
\email{y.takeda@astron.s.u-tokyo.ac.jp}
\author[0000-0001-7154-3756]{Yoshihiro Takeda}
\affiliation{Department of Astronomy, School of Science, The University of Tokyo, 7-3-1 Hongo, Bunkyo-ku, Tokyo, 113-0033, Japan}

\author[0000-0003-3954-4219]{Nobunari Kashikawa}
\affiliation{Department of Astronomy, School of Science, The University of Tokyo, 7-3-1 Hongo, Bunkyo-ku, Tokyo, 113-0033, Japan}
\affiliation{Center for the Early Universe, The University of Tokyo, 7-3-1 Hongo, Bunkyo-ku, Tokyo, 113-0033, Japan}

\author[0000-0002-9453-0381]{Kei Ito}
\affiliation{Cosmic Dawn Center (DAWN), Denmark}
\affiliation{DTU Space, Technical University of Denmark, Elektrovej 327, DK2800 Kgs. Lyngby, Denmark}
\affiliation{Department of Astronomy, School of Science, The University of Tokyo, 7-3-1 Hongo, Bunkyo-ku, Tokyo, 113-0033, Japan}

\author[0000-0001-5394-242X]{Jun Toshikawa}
\affiliation{Nishi-Harima Astronomical Observatory, Center for Astronomy, University of Hyogo, 407-2, Nishigaichi, Sayo, Hyogo 679-5313, Japan}

\author[0000-0002-8857-2905]{Rieko Momose}
\affiliation{Observatories of the Carnegie Institution for Science, 813 Santa Barbara Street, Pasadena, CA 91101, USA}

\author[0000-0002-2205-6115]{Kent Fujiwara}
\affiliation{LY Corporation, Tokyo, Japan}

\author[0000-0002-2725-302X]{Yongming Liang}
\affiliation{Institute for Cosmic Ray Research, The University of Tokyo, Kashiwa, Chiba 277-8582, Japan}

\author[0000-0002-2134-2902]{Rikako Ishimoto}
\affiliation{Department of Astronomy, School of Science, The University of Tokyo, 7-3-1 Hongo, Bunkyo-ku, Tokyo, 113-0033, Japan}

\author[0000-0002-3800-0554]{Takehiro Yoshioka}
\affiliation{Department of Astronomy, School of Science, The University of Tokyo, 7-3-1 Hongo, Bunkyo-ku, Tokyo, 113-0033, Japan}

\author[0009-0007-0864-7094]{Junya Arita}
\affiliation{Department of Astronomy, School of Science, The University of Tokyo, 7-3-1 Hongo, Bunkyo-ku, Tokyo, 113-0033, Japan}

\author[0000-0002-7598-5292]{Mariko Kubo}
\affiliation{Astronomical Institute, Tohoku University, Aoba-ku, Sendai 980-8578, Japan}

\author[0000-0002-0673-0632]{Hisakazu Uchiyama}
\affiliation{National Astronomical Observatory of Japan, 2-21-1 Osawa, Mitaka, Tokyo 181-8588, Japan}



\begin{abstract}
Protoclusters are high-$z$ overdense regions that will evolve into clusters of galaxies by $z=0$, making them ideal for studying galaxy evolution expected to be accelerated by environmental effects. 
However, it has been challenging to identify protoclusters beyond $z=3$ only by photometry due to large redshift uncertainties, hindering statistical study.
To tackle the issue, we develop a new deep-learning-based protocluster detection model, PCFNet, which considers a protocluster as a point cloud.
To detect protoclusters at $z\sim4$ using only optical broad-band photometry, we train and evaluate PCFNet with mock $g$-dropout galaxies based on the N-body simulation with the semi-analytic model. 
We use the sky distribution, $i$-band magnitude, $(g-i)$ color, and the redshift probability density function surrounding a target galaxy on the sky. 
PCFNet achieves to detect five times more protocluster member candidates while maintaining high purity (recall $=7.5\pm0.2$\%, precision $=44\pm1$\%) than conventional methods. 
Moreover, PCFNet is able to detect more progenitors ($M_\mathrm{halo}^{z=0}=10^{14-14.5}\,M_\odot$) that are less massive than supermassive clusters like the Coma cluster. 
We apply PCFNet to the observational photometric dataset of the HSC-SSP Deep/UltraDeep layer ($\sim17\mathrm{\,deg^2}$) and detect $121$ protocluster candidates at $z\sim4$. 
We find the rest-UV luminosities of our protocluster member candidates are brighter than those of field galaxies, which is consistent with previous studies. 
Additionally, the quenching of satellite galaxies depends on both the core galaxy's halo mass at $z\sim4$ and accumulated mass until $z=0$ in the simulation.
PCFNet is very flexible and can find protoclusters at other redshifts or in future extensive surveys by Euclid, LSST, and Roman.
\end{abstract}

\keywords{Protoclusters (1297) --- Lyman-break galaxies (979) --- Galaxy environments (2029) --- High-redshift galaxy clusters (2007)}


\section{Introduction} \label{sec:intro}

There remains an ever-increasing interest in the role of overdense regions as clusters for galaxy formation and evolution.
In the local universe, it is well-known that the properties of cluster and field galaxies are different, such as morphology \citep{Dressler1980, Postman1984}, star formation rates (SFR) \citep{Lewis2002}, and age \citep{Thomas2005}.
Several candidates of characteristic phenomena that make such a difference between environments are conceived, such as the frequency of galaxy mergers \citep{Gottlober2001, Fakhouri2009} and the inflow of cold gas \citep{Kerevs2005, Ocvirk2008}, which trigger and quench star formation, activate the active galactic nuclei (AGNs), and accelerate the growth of black holes \citep{DiMatteo2005,Hopkins2008,Koss2012}.
However, it remains controversial what the major environmental effects are \citep[e.g.][]{Kocevski2012,Mechtley2016,Shah2020}, and further understanding of the properties of galaxies and the role of environmental effects in overdense regions is required.

In the last decades, much work has been done on protoclusters, progenitors of clusters at high-$z$, to understand the origin of the environmental difference. 
\citet{Chiang2017} claimed that protoclusters formed at early epoch $z\sim5\text{-}10$ enhance the star formation of the member galaxies at $z\sim1.5\text{-}5$, and the protocluster core mature with rapidly quenching.
\citet{Muldrew2018} found through their simulations that field galaxies obtain 45\% of their stellar mass at $z=0$, while protocluster galaxies achieve 80\% of their mass by $z=1.4$. 
They thus predicted that the star formation is almost quenched before evolving from protocluster to cluster.
On the other hand, a recent study with cosmological hydrodynamic simulation has reported different star formation rates and phase-transition redshift of protoclusters \citep{Lim2024}.
Observational studies also have shown that the quenching in the protocluster is already progressing at  $1<z<1.5$ \citep{vanderBurg2020, Old2020}, and while high SFR and stellar mass are confirmed for some spectroscopically identified protoclusters between $2<z<3$ \citep{Hatch2011, Koyama2013, Cooke2015, Strazzullo2018, Shimakawa2018, WangT2018}, some protoclusters of massive quiescent galaxies at $2<z<4$ are also found by recent spectroscopic observation \citep{McConachie2022, Kubo2022, Kalita2022, Ito2023, Tanaka2023}.
Thus, there is still room for discussion on the properties of protocluster, especially at high redshift.
Beyond $z\gtrsim4$, several protoclusters are detected by various methods and identified with follow-up spectroscopic observation \citep[e.g.,][]{Capak2011, Walter2012, Toshikawa2012, Trenti2012, Ishigaki2016, Oteo2018, Miller2018, Harikane2019, Calvi2021, Hu2021, Castellano2022, Sillassen2022, Larson2022, Endsley2022, Wang2024}, with a recent increase in the detections at high redshifts by JWST \citep[e.g.][]{Hashimoto2023, Morishita2023, Helton2024}.
Moreover, there are recent attempts to search protoclusters with combined spectroscopic and photometric data by the Voronoi Monte-Carlo algorithm to obtain six massive protoclusters at $z\sim2\mathchar`-5$ \citep[e.g.,][]{Shah2024}.
However, the number of identified protoclusters is still small \citep{Overzier2019} because high-$z$ protoclusters are extremely rare (e.g., $\sim10^{-5}\mathrm{\, Mpc^{-3}}\,\mathrm{at}\,z\sim4$) in space density and sufficiently wide and deep spectroscopic surveys have not been carried out.
In addition, the detection methods differ among them, making a bias-free statistical discussion of this heterogeneous sample difficult.

Fortunately, the Hyper Suprime-Cam Strategic Survey Program \citep[HSC-SSP;][]{Aihara2018} has been conducted to observe a remarkably wide region with $g$ to $y$-band and enables us to select galaxies at $z\sim4\mathchar`-6$ with the dropout technique \citep{Ono2018}.
Especially, the number of $g$-dropout galaxies at $z\sim4$ is enough to search protoclusters so that they are ideal targets to shed light on high-$z$ protocluster search. \citet{Toshikawa2018} conducted a systematic protocluster search with dropout galaxies at $z\sim4$ to detect 179 unique protocluster candidates over an area of 121 deg$^2$ from the HSC-SSP Wide layer.
The candidates of protocluster galaxies tend to have bright-end excess in their UV luminosity function and excess emission in mid-IR, which implies that star formation enhances in protocluster beyond $z>4$ \citep{Ito2020, Kubo2019}. 
Additionally, quasars are distributed to avoid the most overdense regions, \citep{Uchiyama2018, Uchiyama2020}, but the pairs of quasars tend to occur in massive halo \citep{Onoue2018}.
Although these studies provide a foothold for a statistical survey of the high-$z$ protocluster, the detection performance is degraded by the projection effect (completeness$\sim6\mathchar`-13\%$), due to the large uncertainty in the redshifts of dropout galaxies. 
Moreover, the samples are biased towards very massive protoclusters that evolve into the most massive structures, such as the Coma cluster; $M_\mathrm{halo}^{z=0}>10^{15}M_\odot$. 
This bias may lead to extreme characteristics when examining environmental effects.
Theoretical studies \citep[e.g.,][]{Yajima2022} suggest that there may be essentially different physics at work in low-mass haloes ($M_\mathrm{halo}<10^{12.5}M_\odot$) compared to those of high-mass: the star formation in low-mass haloes is suppressed due to the supernova feedback, and a part of the metal-enriched gas can be expelled by the galactic winds, leading to a steep mass dependence of metallicity. 
To verify this, it is necessary to study protoclusters whose halo mass is still $M_\mathrm{halo}<10^{12.5}M_\odot$ at high redshift.
According to dark matter N-body simulation, such low-mass haloes at $z\sim4$ evolve into $M_\mathrm{halo}\sim10^{14}M_\odot$ at $z=0$ on average \citep{Chiang2013}.
Hence, a statistical survey for low-mass protoclusters at $z\sim4$ has the potential clue to obtain a more general picture.

In recent years, advances in information science have led to deep learning approaches being applied in a variety of fields.
As an example applied to the large-scale structure, \citet{Inoue2022} used a Convolutional Neural Network (CNN) to classify the structure, and \citet{Chen2023, Ganeshaiah2023, WuZ2023} used a CNN to solve the density field reconstruction task.
\citet{Anagnostidis2022} used PointNet \citep{Qi2016} to estimate cosmological parameters. \citet{WuF2023} used Graph Neural Network (GNN) to infer galaxy properties from dark matter information.

Here, we present a new model to detect protocluster candidates at $z\sim4$ which evolve into $M_\mathrm{halo}\sim10^{14}M_\odot$ at $z=0$ solely from photometric data by a deep learning approach, PCFNet.
PCFNet is developed based on the PointNet \citep{Qi2016} and DG-CNN \citep{Wang2018}, which handle point cloud data because the distribution of galaxies at $z\sim4$ is relatively sparse (see Sec \ref{ssec:pointcloud}).
Furthermore, the uncertainties of the location of each galaxy along the line-of-sight are very large; therefore, it would be better to adopt the point cloud format to maximize the use of the information.

The contents of this paper are as follows.
In section \ref{sec:data}, we describe the simulation data used in this paper. 
In addition, we explain the selection method of $g$-dropout galaxies and the definition of protoclusters. 
Section \ref{sec:method} consists of the details of our method. 
We demonstrate the results of our method's performance in section \ref{sec:result} and application to observational data in section \ref{sec:applyobs}. 
In section \ref{sec:discussion}, we dig into the properties of protoclusters from simulational and observational points of view.
Lastly, we summarize our contributions in section \ref{sec:conclusion}.

We assume the following cosmological parameters estimated by the Planck1 mission results \citep{Planck2014}: $\Omega_m = 0.315, \Omega_\Lambda = 0.685, H_0 = 67.3 \mathrm{\,km\,s^{-1}\,Mpc^{-1}}$ following \citet{Araya-Araya2021}, and use AB magnitude system \citep{Oke1983}.

\section{Data}
\label{sec:data}

The simulation data are used to train and evaluate the deep learning model and to predict the nature of the protocluster member galaxies. 
It would be preferable to use real data from already observed protoclusters as training data. However, the number of protoclusters at $z \sim 4$, in which most of the member galaxies are spectroscopically identified, is very limited and cannot be used as training data.
We use a semi-analytic model, PCcone \citep{Araya-Araya2021}, as an alternative.
PCcone is the best choice for training data because it reproduces the observed data, especially for the HSC-SSP S20A photometric data \citep{Aihara2022}, as described in section \ref{ssec:lbg}. 
It should be noted that the semi-analytic model does not perfectly reproduce the actual universe \citep{Lim2021}, and it is inevitable that the predictions of a model trained on this basis are influenced by the model.
However, PCcone is the only light cone model that has sufficient volume for studying protoclusters and reproduces the distribution and photometry of galaxies beyond $z\sim4$.

\subsection{Simualtion Data -- PCcone}
\label{ssec:simdata}

To train and evaluate our deep learning model, we use PCcone, which is the light-cone model generated based on the Millennium Simulation \citep{Springel2006} and semi-analytic model, \texttt{L-GALAXIES} \citep{Henriques2015} to simulate a photometric redshift survey of protoclusters\footnote{PCcone is adjusted to place protoclusters at the specified redshift $(z=1.0,\,1.5,\,2.0,\,2.5,\,3.0)$.
However, since we mainly use the PCcone beyond $z=3$, we do not use the peculiarity of the protocluster arrangement.
The effect of this placement singularity on the results is small.}.
The Millennium simulation has a volume of $L = 480.279 \mathrm{\, Mpc\, }h^{-1}$, which is large enough to contain sufficient protoclusters, and a dark matter particle mass of $m_p = 9.6 \times 10^8\, M_\odot\mathrm{\, }h^{-1}$.
The PCcone provides photometric predictions to reproduce the observed conditions of HSC-SSP, Deep Canada–France–Hawaii Telescope Legacy Survey (CFHTLS)\footnote{\url{https://www.cfht.hawaii.edu/Science/CFHLS/}}, and Legacy Survey of Space and Time \citep[LSST;][]{Ivezic2019}, which can be manually changed.
We set a photometric error corresponding to the $5\sigma$ depths shown in Table \ref{table:5sigma}, which is based on the HSC-SSP S20A Deep and UltraDeep observing conditions \citep{Aihara2022}.
Since Deep and UltraDeep have different photometric errors, we prepare two simulation data generated for Deep and UltraDeep.
In addition, as with the setting of the limiting magnitude in the observational data, the data are restricted to bright objects with $i<26$.

We use 20 light cones randomly selected from PCcone, each with a field of view of radius 1$^\circ$,
and separate 15 as training data, one as validation data, and four as evaluation data.
We note that the same galaxies may appear in training and validation/evaluation data because light cones are made from the same simulation boxes on Millennium Simulation. 
However, the input information (e.g., coordinates or magnitudes; see Sec \ref{ssec:ourmodel}) varies because the direction of the line of sight and the resampled magnitudes differ, and the possibility of critical overtraining or leakage is slight. 
Unique training and evaluation samples adjusted to Deep (UltraDeep) data on Millennium Simulation are 309646 (292421) and 82245 (77986): the number of galaxies in the evaluation data that originate from the same galaxies in training data is 11264 (10345).

\begin{table}[t]
    \centering
    \begin{tabular}{cccccc}
        \hline
         & $g$ & $r$ & $i$ & $z$ & $y$ \\
        \hline
        \hline
        Deep & 27.354& 26.974& 26.752& 26.246& 25.316\\
        UltraDeep & 28.223 &27.869 &27.850 &27.313 &26.343\\
        \hline
    \end{tabular}
    \caption{$5\sigma$ limiting magnitude for the PCcone.}
    \label{table:5sigma}
\end{table}

\subsection{Dropout Galaxies}
\label{ssec:lbg}

We select the dropout galaxies at $z\sim4$ based on the Lyman break technique \citep[e.g.,][]{Steidel1996}.
Our criteria are the following, which are determined with reference to the \citet{Ono2018}:
\begin{equation}
     \begin{split}
     &(g-r > 1.0)\\
     \land\quad &(-1.0 < r-i < 1.0)\\
     \land\quad &(1.5(r-i) < g - r -0.8)\label{eq:gdropout}.
     \end{split}
\end{equation}
If the $r$-band magnitude is greater than the $2\sigma$ limiting magnitude ($m_{\mathrm{lim,2}\sigma}$), i.e., there is no detection in $r$-band, we use the $2\sigma$ limiting magnitude instead of the observed magnitude to avoid the uncertain magnitudes with insufficient significance affecting the selection.
Additionally, we only use the galaxies that have $g<m_{\mathrm{lim,2}\sigma}$ to exclude the dark galaxies for which redshift estimation is difficult (see Sec \ref{sssec:mdn}).

\begin{figure*}[th!]
    \centering
    \includegraphics[width=\textwidth]{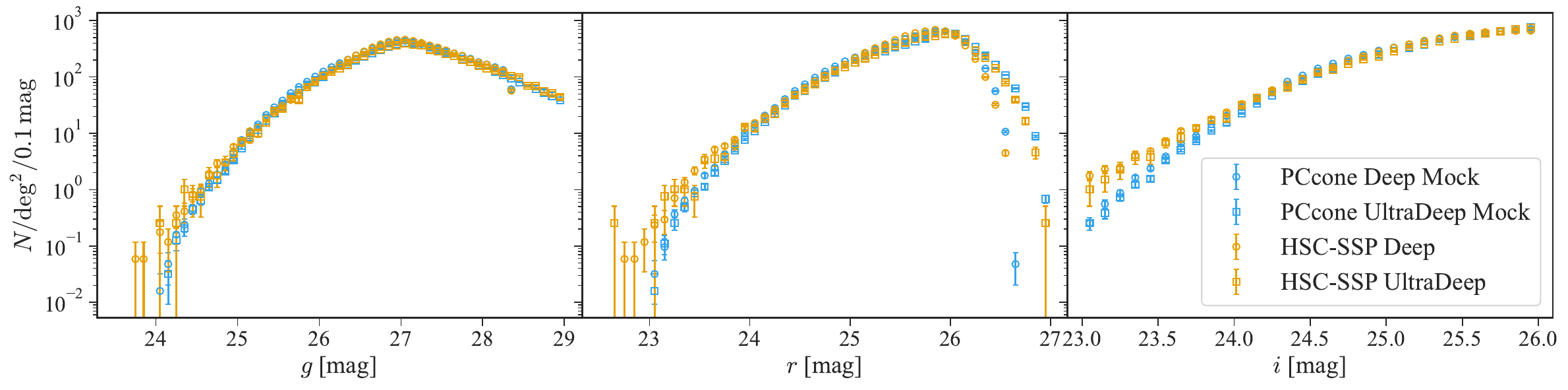}
    \caption{The $g,\,r,\,i$-bands magnitude distribution of $g$-dropout galaxies at $z\sim4$ in the simulation (PCcone; light blue) and the observational (HSC-SSP; yellow) data. The circle and square represent the Deep and UltraDeep layers, respectively. The error bars of each point indicate Poisson errors.}
    \label{fig:lbgnd}
\end{figure*}

We select 421634 and 397762 $g$-dropouts from the PCcone adapted to the HSC-SSP Deep and UltraDeep layers, respectively.
Figure \ref{fig:lbgnd} shows $g,\,r,\,i$-band magnitude distribution of $g$-dropouts over the whole redshift range in PCcone.
In the figure, the observational data from HSC-SSP are also shown for comparison.
The details of the selection for the observational data are described in section \ref{ssec:obsdata}.
The faint end of each distribution between PCcone and HSC-SSP is consistent, while the brighter end of HSC-SSP is a little greater than that of PCcone.
The difference might be due to the contamination of stars \citep{Capak2004} or magnitude offsets that are not fully reproduced by the simulation \citep{Araya-Araya2021}.
However, they are few compared to the total; therefore, the effect when applied to observed data is negligible.
We also check the distribution of magnitudes and colors as a function of redshifts and confirm there is no abnormality such as an artificial gap due to the transition of snapshot in the simulation (comparisons not shown here).
Note that low redshift contaminants are known to be included in the dropout selection \citep[e.g.,][]{Ono2018}. 
The percentage of the outliners in our selection is 2.9\% from simulation data.
In the following, we use the term $g$-dropout galaxies at $z\sim4$ for readability, but it does not mean the $g$-dropout galaxies from which all low-$z$ contamination has been removed.

\begin{figure}[tbh!]
    \centering
    \includegraphics[width=0.45\textwidth]{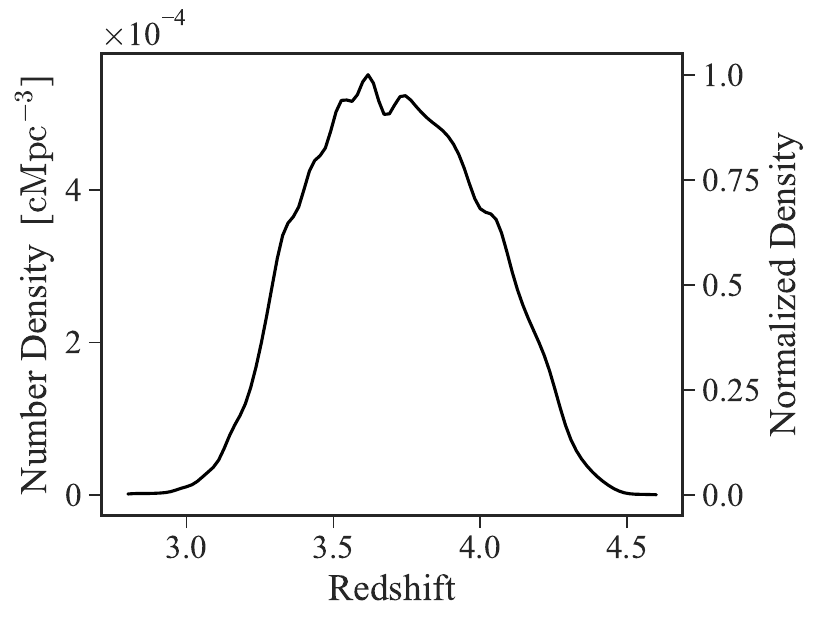}
    \caption{Number density as a function of redshift for $g$-dropout galaxies in the simulation data. The right vertical axis represents the normalized density.}
    \label{fig:simgdropoutcomp}
\end{figure}

We stack all the light cones of PCcone and calculate the mean number density of $g$-dropout galaxies per $40 \mathrm{\,cMpc\,}h^{-1}$ in line of sight.
Figure \ref{fig:simgdropoutcomp} shows the number density as a function of redshift interpolated by a cubic function for $g$-dropout galaxies in PCcone.

\subsection{Protocluster Definition}
\label{ssec:pc_define}
\begin{figure}[t!]
    \centering
    \includegraphics[width=0.45\textwidth]{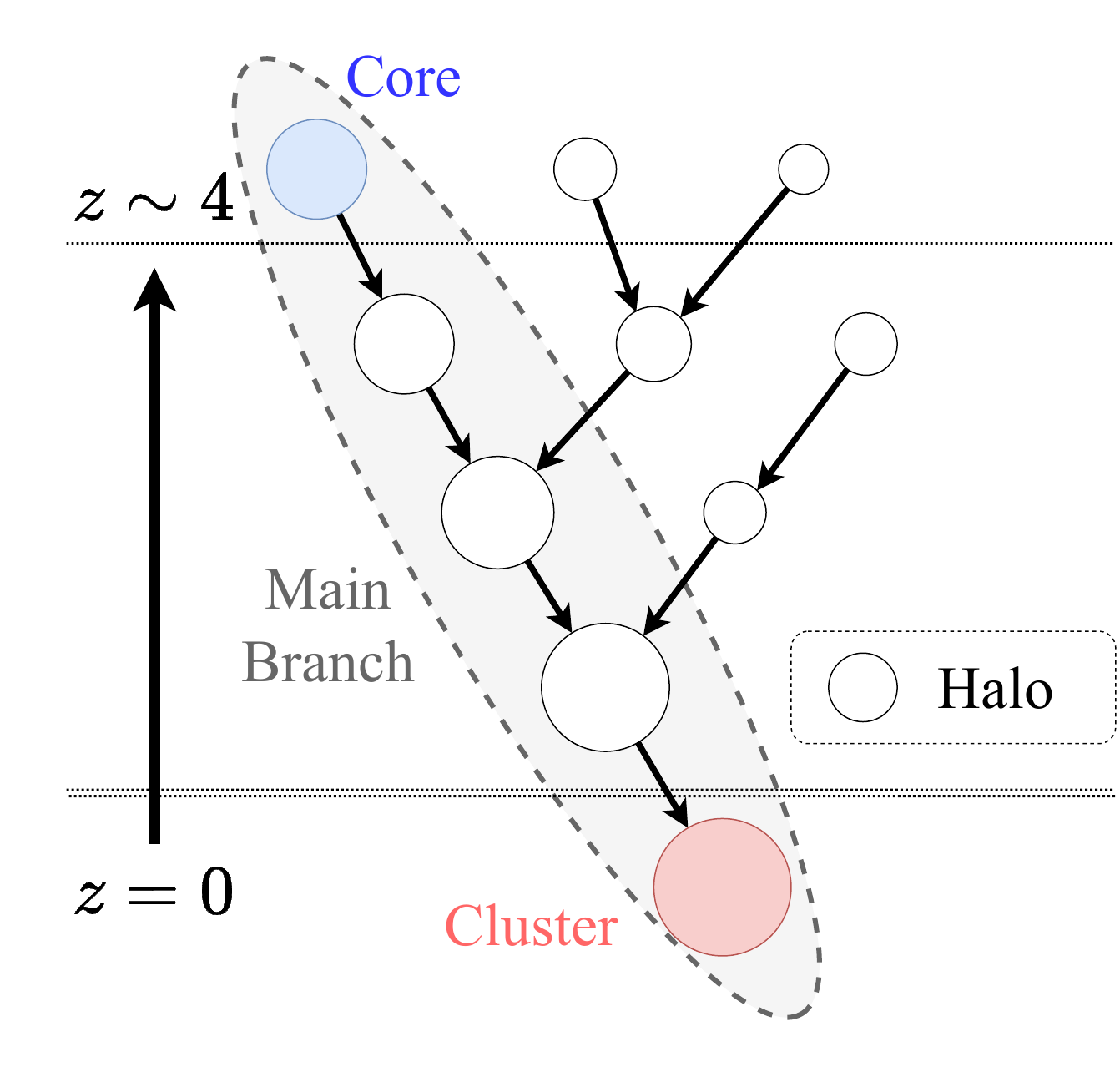}
    \caption{Diagram of a merger tree. Each circle represents a halo merging from top to bottom. Larger circles indicate more massive haloes. The red circle shows the cluster that is the root of the merger tree, and the blue circle shows the halo containing the core galaxy at $z = 4$. The gray area surrounded by the dashed line is the main branch in the merger tree, and the core galaxies are always included in the main branch.}
    \label{fig:mergertree}
\end{figure}

Although the definition of a protocluster varies in previous studies \citep[e.g.][]{Chiang2013, Muldrew2015}, in this study, we define a protocluster based on the merger trees as follows (Figure \ref{fig:mergertree}). We select galaxies with central halo masses exceeding $M_\mathrm{halo}^{z=0}=10^{14}\,{M_\odot}\,h^{-1}$ at $z=0$ from Millennium Simulation.
We use $m_{\mathrm{tophat}}$, the mass within the halo radius with an excess density equal to the threshold of the spherical collapse model, as the halo mass.
By retracing each galaxy's merger tree \citep{Lemson2006}, we ascertain their main progenitors at $z=4$ (hereafter called core galaxies)\footnote{We define the progenitors with the largest halo mass at $z\sim4$ by tracing the main branches of the merger tree.
The main progenitors are usually identified with the most massive progenitors. However, there is a possibility deviated from the suitable series by momentary noise.
To prevent this, the main progenitors are recursively redefined to follow the history of the most massive progenitors for a long time. For more information, see \citet{Lucia2007}}.

We label galaxies within $R=5.5$ cMpc radius of the core galaxy as member galaxies.
The radius is set up based on the mean effective radius ($R_e$) of the protoclusters at $z\sim4$ whose members are defined as all progenitors of the cluster, $M_\mathrm{halo}^{z=0}>10^{14}M_\odot h^{-1}$, derived from the merger tree.
The effective radius is defined as \citet{Chiang2013} Eq. 4:
\begin{equation}
    R_e=\sqrt{\dfrac{1}{M_\mathrm{halo}}\sum_i m_i(\mathbf{x}_i-\mathbf{x}_c) },
\end{equation}
where $m_i$ and $\mathbf{x}_i$ are the halo mass and position of each galaxy, and $\mathbf{x}_c$ is the center of protocluster members, respectively.
The member galaxies defined in our way make up 95\% of the galaxies that merge by $z = 0$ according to the merger tree.
The above procedure enables us to flag only those galaxies significantly concentrated at $z=4$.

We enumerate the member galaxies based on the central halo at $z = 0$ and a proto-cluster is defined as a region with a certain number of members.
It should be noticed that the number of member galaxies depends on the dropout selection (see Figure \ref{fig:simgdropoutcomp}).
We divide the number by the normalized number density of $g$-dropout galaxies at the redshift of the core galaxy to correct the effect:
\begin{equation}
    N_{\mathrm{mem,c}}=N_{\mathrm{mem}}/\gamma\,,
\end{equation}
where $N_{\mathrm{mem,c}}$ is the corrected number of member galaxies, $N_{\mathrm{mem}}$ is the number of member galaxies, and $\gamma$ is the correction rate, which is the normalized number density of the $g$-dropout galaxies at the redshift of the core galaxy.
To avoid overcorrection at very low rates, the redshift range of less than 50\% overcorrection rate is excluded from the protocluster search.
Then, a protocluster is defined as a group whose corrected number of member galaxies is greater than $N_{\mathrm{th}}=5$.
We note that changing this threshold into $N_\mathrm{th}=3\mathchar`-7$ does not affect the following results.

In the simulation data, the number of protoclusters is 4262 and 4345  at $3.28<z<4.14$, and the total number of member galaxies belonging to a protocluster is 59116 and 54705 for the Deep and UltraDeep layers, respectively.
The maximum halo mass of the protoclusters is $2.9\times10^{15}M_\odot$.

\section{Methods}
\label{sec:method}

\subsection{Deep learning with point clouds}
\label{ssec:pointcloud}
We use a point cloud-based deep learning model to deal with the galaxy distribution. 
A point cloud is a set of points represented by vectors, and the vectors are generally described in a three-dimensional space with additional information. In the field of information science, point clouds mainly represent the surface of an object obtained by devices such as Light Detection And Ranging (LiDAR) sensors \citep[e.g.,][]{Geiger2012}. 
The merit of point clouds is the ability to represent sparse data more efficiently than voxels or meshes.
In protocluster detection, it is difficult to integrate the information of each galaxy into voxels, and the distribution of galaxies is very sparse compared to the size of the universe.
Therefore, it is desirable to analyze the galaxy distribution by representing them as a point cloud.

PointNet \citep{Qi2016} is a point cloud-based deep learning model, which is the foundation of recent point cloud analysis with deep learning.
PointNet achieved higher performance than a conventional 3D fully convolutional Network, which is processed by voxelizing and convolutional neural networks, by adopting a network structure that considers the symmetry of the order-invariance and transform-invariance of point clouds.
DG-CNN \citep{Wang2018} is one of the successor models of PointNet, which takes localization in the network by dynamically computed $k$-neighbor graphs in each network layer.

\subsection{Our approach}
\label{ssec:ourmodel}
We propose a three-step approach to detect protocluster candidates: First (section \ref{sssec:mdn}), we preprocess the input galaxy data with the Mixture Density Network \citep{Bishop1994} to estimate the redshift probability density function (PDF) with three optical band magnitudes ($g,\,r,\,i$). Secondly (section \ref{sssec:pcfnet}), we use the new deep-learning model, PCFNet, to predict the probability that the target galaxy is a protocluster member. Finally (section \ref{sssec:grouping}), we group the protocluster member candidates into protocluster candidates with persistent homology. 

It is worth noting the limitations or assumptions of the deep learning model in our approach. 
Simulation data to use in training steps should have well-known and realistic photometric properties (e.g., limiting magnitude, see section \ref{ssec:lbg}).
It is also important that the dropout band is deep enough and that the number of detections is sufficient to identify the protocluster structure (e.g., $\gtrsim 0.1 \mathrm{cMpc^{-3}}$ at $z\sim4$). 
PCcone satisfies the first prerequisite since its magnitude limit is customizable to match the survey program used (see section \ref{ssec:simdata} for more details).
The $g$-dropout sample used in this study has the largest number of galaxies among the HSC-SSP dropout galaxy samples, and it fits the precondition well.

\subsubsection{Mixture Density Network}
\label{sssec:mdn}
To input galaxy features into PCFNet, we first estimate the PDF of line-of-sight distances of the galaxies from the $g,\,r,\,i$ band magnitudes by the Mixture Density Network \citep[MDN;][]{Bishop1994}. Several previous studies use the same method to infer the redshift \citep[e.g.,][]{DIsanto2018, Ansari2021}. 
The merit of MDN is obtaining the output as a PDF which is a mixture of several Gaussian distributions.
Since similar spectral continuum breaks, such as the Lyman break or Balmer break, appear on the galaxy spectra, estimated redshift distribution from a few bands generally has two or more peaks. 
For a stable estimate of the redshift of a point cloud, therefore, it is better to consider the entire PDF, including multiple peaks, than a single best value.

We limit to three bands on the assumption that in the future PCFNet will be applied to HSC-SSP Wide, which has an ultra-wide field of view ($\sim900\mathrm{\,deg^2}$) suitable for detecting PCs, in addition to the unavailability of accurate photo-$z$.
We confirm that there is a small increase in accuracy even when $z,\,y$ bands are added to the input data. 

MDN is composed of a three-layer neural network and outputs three values corresponding to mean $\bm{\mu}(\textbf{x})$, variance $\bm{\sigma}(\textbf{x})$, and mixing coefficient weights $\bm{\omega}(\textbf{x})$ for the input $\textbf{x}$.
MDN approximates the PDF of the target variable $y$ by the mixed Gaussian distribution $P(y|\textbf{x})$ expressed by the following equation:
\begin{equation}
    P(y|\textbf{x}) = \sum_{k=1}^{K} \omega_k(\textbf{x}) \phi(y|\mu_k(\textbf{x}), \sigma_k^2(\textbf{x})),
\end{equation}
where
\begin{equation}
\phi(y|\mu_k(\textbf{x}), \sigma_k^2(\textbf{x}))=\frac{1}{\sqrt{2\pi\sigma_k^2(\textbf{x})}}\exp\left(-\frac{(y-\mu_k(\textbf{x}))^2}{2\sigma_k^2(\textbf{x})}\right) .   
\end{equation}
The parameters of the mixture Gaussian distribution can be written as the output values of the neural network $\mathbf{z}^\mu,\,\mathbf{z}^\sigma,\,\mathbf{z}^\omega$, assuming a uniform distribution as the prior. They can be written in the following form:
\begin{align}
    \bm{\mu}(\textbf{x}) &= \mathbf{z}^\mu\\
    \bm{\sigma}(\textbf{x}) &= \exp{(\mathbf{z}^\sigma)} \\
    \bm{\omega}(\textbf{x}) &= \frac{\mathbf{z}^\omega}{\sum_{k=1}^{K} {z}^\omega_k} .
\end{align}
We approximate the line-of-sight comoving distance by mixing $K=3$ Gaussians. The accuracy did not change when the number of Gaussian mixtures was increased to $K=5$. The middle layer has 20 dimensions.

The MDN is trained using 15 light cones as training data. The learning is performed for 50,000 epochs using the line-of-sight comoving distance of the galaxies as ground truth.
The learning rate to update the weights of MDN was set to $\mathrm{lr}=0.001$, and the optimization algorithm was Adam \citep{Kingma2014}.

After training the MDN, we apply it to all of the training, validation, and evaluation data and calculate the probability distribution of the distance for each galaxy. The resulting distribution is expressed in nine parameters, three each for the mean, variance, and mixing coefficient.

\subsubsection{MDN performance}
To evaluate the accuracy of MDN, we obtain the peak values, $z_{\mathrm{best}}$ from the probability density functions estimated by MDN.
Then, we calculate the root mean square error, $\mathrm{RMS}(\delta z)_{\mathrm{all}}$, where $\delta z = (z_{\mathrm{best}}-z_{\mathrm{true}})/(1+z_{\mathrm{true}})$ is the error and $z_{\mathrm{true}}$ is the true redshift in the simulated data.
Furthermore, we consider the sources with $|\delta z|>\mathrm{RMS}(\delta z)_{\mathrm{all}}$ as outliers and calculate the mean $\mean{\delta z}$ and root mean square $\mathrm{RMS}(\delta z)$ excluding them.
A comparison of the performance for the evaluation data with another redshift estimator (\textsc{eazy}; \citealt{Brammer2008}) is shown in Table \ref{table:photoz}. 
We run \textsc{eazy} with the default templates created by the Flexible Stellar Population Synthesis \citep[FSPS; ][]{Conroy2009, Conroy2010} code for a range of redshifts $0.01<z<5$ in a step of $\varDelta z=0.005$.
As with MDN, only $g, r, i$-bands are inputted into \textsc{eazy}.
As seen in Table \ref{table:photoz}, MDN has smaller errors and systematic offsets than \textsc{eazy}, and the RMS of MDN is also consistent in that of the photo-$z$ of HSC-SSP Public Data Release (PDR) 2 \citep{Nishizawa2020}.

\begin{table}
    \centering
    \begin{tabular}{ccc}\hline
     &  $\mean{\delta z}$&$\mathrm{RMS}(\delta z)$\\\hline\hline
     MDN& $-0.001$ & $0.045$\\
     \textsc{eazy}& $0.020$ & $0.068$\\\hline
    \end{tabular}
    \caption{The mean and root square error for estimated redshift by MDN and \textsc{eazy}.}
    \label{table:photoz}
\end{table}

\begin{figure}
    \centering
    \includegraphics[width=0.48\textwidth]{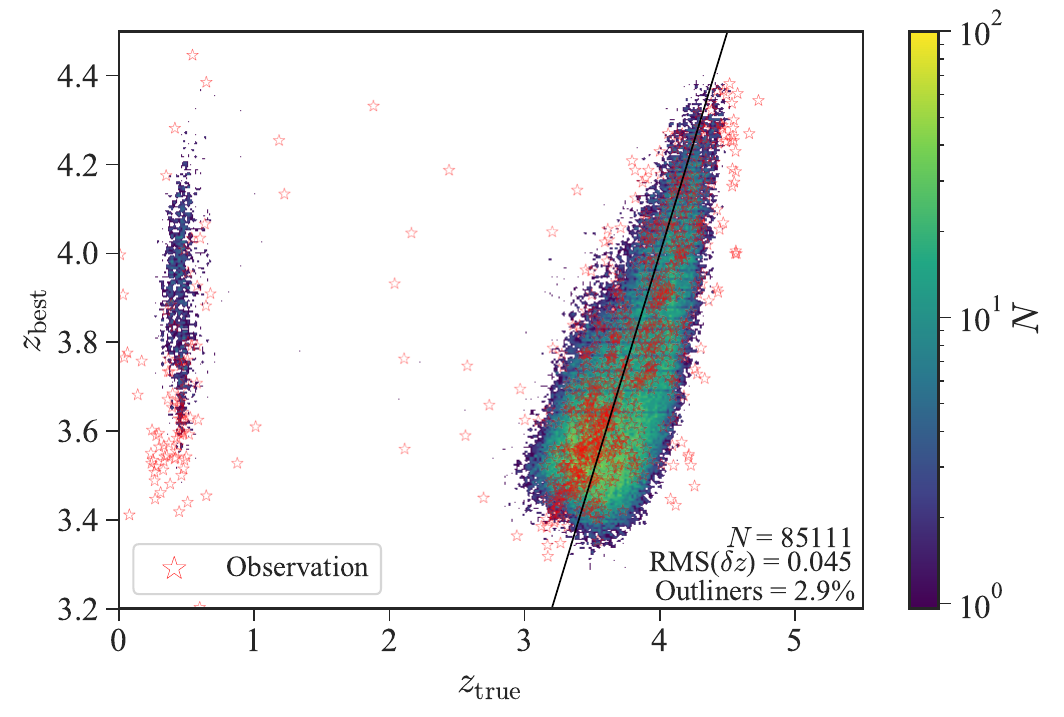}
    \caption{The MDN predictions $z_\mathrm{best}$ vs. true distances $z_\mathrm{true}$ in the simulation data. The color of each point represents the number of galaxies. The $g$-dropout galaxies in the HSC-SSP UltraDeep layer with known spectroscopic redshifts are marked with red stars.}
    \label{fig:mdn_precision}
\end{figure}

The estimated redshift $z_{\mathrm{best}}$ compared with the true redshift $z_{\mathrm{true}}$ is shown in Figure \ref{fig:mdn_precision}.
We also show the results applied to sources with known spectroscopic redshifts in the HSC-SSP Deep layer\footnote{Since galaxies with known spectroscopic redshifts in the Deep layer are few, the combined results of galaxies in the UltraDeep layer are shown here. 
We find no significant difference in MDN performance between the Deep and UltraDeep regions.
} retrieved from HSC-SSP \textit{specz} table.
This table is composed of 
zCOSMOS DR3 \citep{Lilly2009}, 
UDSz \citep{Bradshaw2013,McLure2013},
3D-HST \citep{Skelton2014,Momcheva2016},
FMOS-COSMOS \citep{Silverman2015},
VVDS \citep{LeFevre2013},
VIPERS PDR1 \citep{Garilli2014},
SDSS DR16 \citep{Ahumada2020},
SDSS QSO DR14 \citep{Paris2018},
GAMA DR2 \citep{Liske2015},
WiggleZ DR1 \citep{Drinkwater2010},
DEEP2 DR4 \citep{Davis2003,Newman2013},
DEEP3 \citep{Cooper2011,Cooper2012},
PRIMUS DR1 \citep{Coil2011,Cool2013},
2dFGRS \citep{Colless2003},
6dFGRS \citep{Jones2004,Jones2009},
C3R2 DR2 \citep{Masters2017,Masters2019},
DEIMOS 10k sample \citep{Hasinger2018}, and
LEGA-C DR2 \citep{Straatman2018}:
For the details of the selection, see \citet{Aihara2022}.
Note that the table has a selection bias due to restrictions for secure samples.
The MDN estimation results are scattered around the line $z_{\mathrm{true}}=z_{\mathrm{best}}$, and there is no systematic bias in the prediction results except for the low redshift contamination.

\subsubsection{PCFNet}
\label{sssec:pcfnet}
\begin{figure*}[tbh!]
\centering
\includegraphics[width=0.9\textwidth]{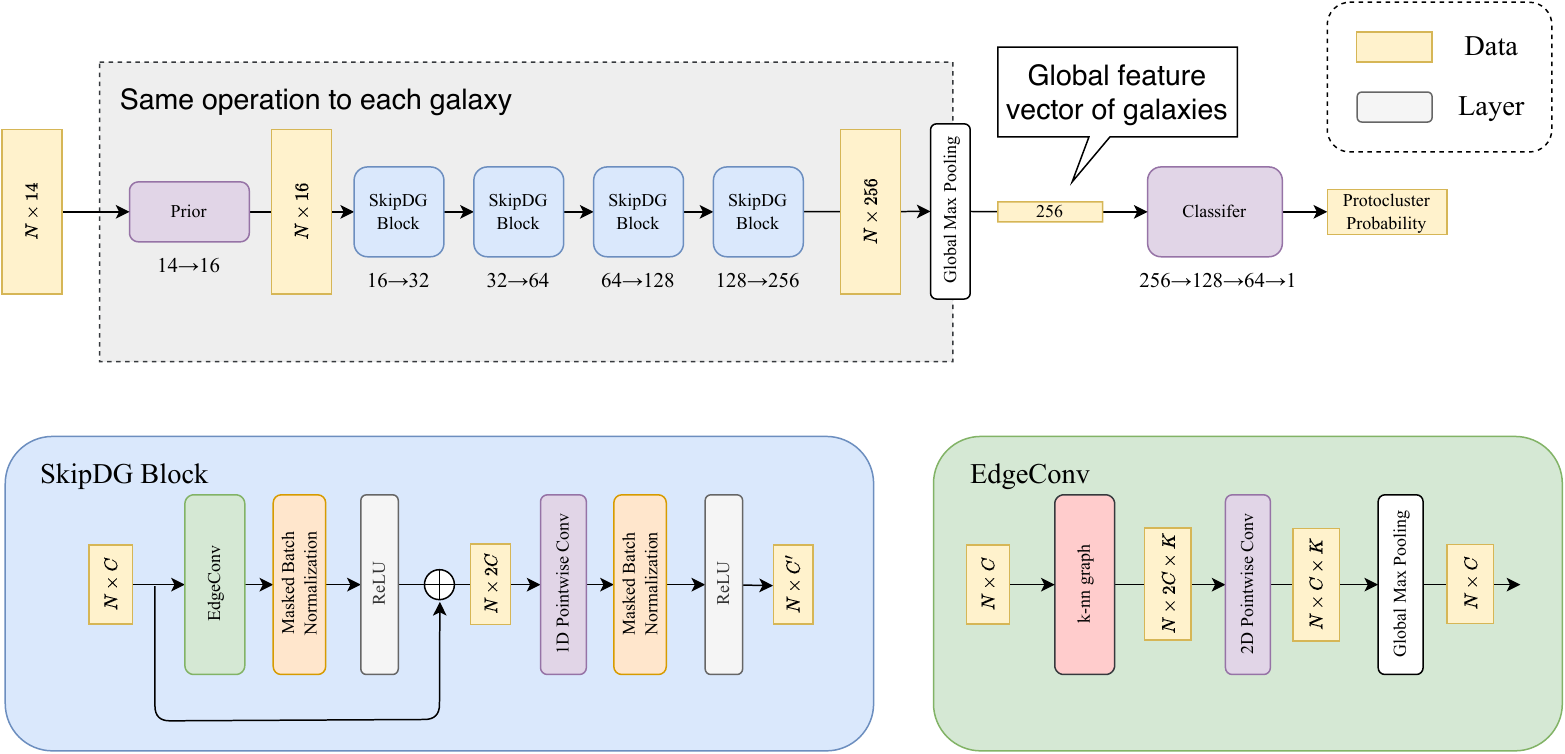}
\caption{The architecture of PCFNet.
The round squares represent each layer, and the numbers at the bottom show the input and output dimensions of the features.
The squares represent the data at each stage, and the dimensions of the matrix are written inside.
Prior is a single-layer shared-weight MLP, and Classifier is a three-layer MLP.
}
\label{fig:pcfnet}
\end{figure*}

We develop a new point cloud-based deep learning model as a protocluster finder, PCFNet, inspired by PointNet and DG-CNN.
PCFNet comprises the following four parts: a prior layer, skipDG blocks, a global max pooling layer, and a classifier (Figure \ref{fig:pcfnet}).
In the prior layer, the input features of each galaxy are expanded to 16 dimensions, which is a combination of a pointwise convolutional layer, batch normalization, and Rectified Linear Unit (ReLU, \citealt{Glorot2011}) as an activation function.
The first half of the skipDG block consists of EdgeConv \citep{Wang2018}, batch normalization, and ReLU, which output is then combined with the original input of the skipDG block.
EdgeConv plays a role in capturing the local features of the point cloud:
EdgeConv first constructs a local neighborhood graph, $\mathcal{G}^l$, from point clouds, $\mathcal{P}^l\subseteq\mathbb{R}^{N\times C_l}$, which $l$ is an index of EdgeConv, $C_l$ is the dimension of input matrix of the $l$th EdgeConv, and $N$ is the number of points.
Then, $K$ neighborhoods, $\mathbf{x}^l_{i_k}\ (1\leq k\leq K)$ are selected by each point, $\mathbf{x}^l_i \ (1\leq i\leq N)$, and new feature vectors, $\mathbf{x}^l_{(i,k)}\in \mathbb{R}^{2C_l}$, are made as the following:
\begin{equation}
    \mathbf{x}^l_{(i,k)}=\left(
    \begin{matrix}
        \mathbf{x}^l_{i}\\
        \mathbf{x}^l_{i_k} - \mathbf{x}^l_{i}
    \end{matrix}
    \right).
\end{equation}
After that, the pointwise convolution, $h_\Theta$, and global pooling among the neighborhood features are applied to the vectors:
\begin{align}
    h_\Theta\left(\mathbf{x}^l_{(i,k)}\right) &= \mathrm{ReLU}\left(\Theta \mathbf{x}^l_{(i,k)}\right)\\
    \mathbf{x}^{l+1}_{i} &= \max_{1\leq k\leq K} \left\{h_\Theta\left(\mathbf{x}^l_{(i,k)}\right)\right\},
\end{align}
where $\Theta\in\mathbb{R}^{C_{l+1}\times2C_l}$ is the shared weights of the pointwise convolution.
The padding features are replaced by a small enough value to mask meaningless values when global pooling.
The second half of the skipDG block is the network of sequences; pointwise convolution layer, batch normalization, and ReLU.
Four stacked skipDG blocks enhanced the features to $16\rightarrow32\rightarrow64\rightarrow128\rightarrow256$.
The global max pooling layer compresses the pointcloud features into a 256-dimensional vector.
The role of the classifier is to reduce the dimensions of the vector ($256\rightarrow128\rightarrow64\rightarrow1$) to output a probability that the target galaxy is a protocluster member.
The classifier consists of several layers in the order of the dense layer, batch normalization, and ReLU. A dropout layer is sandwiched between the second and third classifier layers, and dropout is performed during training with a probability of $p=0.3$. The sigmoid function is used as the activation function for the final layer.

We input the following values of all galaxies (say $N$ from the following) in a radius of $5'$ around the target galaxy; the sky coordinates, $d_{\mathrm{samp}}$, $i$ band magnitude, $(g-i)$, and the nine parameters of the probability distribution described above.
We note that the radius is set as the diameter of a protocluster at $z\sim4$ to cover almost all members \citep{Chiang2013}.
It is also worth mentioning that $N$ varies depending on the target galaxy. 
To enable parallel processing, the dimensions of the matrix are set to the maximal number of neighboring galaxies in the dataset, and the matrix is padded for those with fewer neighbors.
The padding spaces in the batch normalization and global max-pooling layers are excluded so that the padding does not affect the result. We train PCFNet using the flag of protocluster member galaxies as the class label. The cross-entropy error is used as the loss function:

\begin{gather}
  E=-\dfrac{1}{N_b}\sum_{i=1}^{N_b}\left(y_i\log{p_i} + \left(1-y_i\right)\log{\left(1-p_i\right)}\right),
\end{gather}
where $N_b$ is the size of the mini-batch, $\mathbf{y}=\left(y_1, \dots,y_{N_{b}}\right)^\top$ is the one-hot vector for being a protocluster member galaxy, and $\mathbf{p}=\left(p_1, \dots, p_{N_{b}}\right)^\top$ is the output of PCFNet. 
The Adam \citep{Kingma2014} optimization algorithm is used with initial learning rate $\mathrm{lr}=0.001$, momentum 0.9, and batch size 512.
The learning rate is multiplied by 0.5 if the cross-entropy error does not decrease by more than 0.9 as a ratio over three steps.
We use early stopping if there is no improvement for 15 epochs.
PCFNet is trained with a Tesla V100 32GB GPU.

\subsubsection{Grouping as protocluster}
\label{sssec:grouping}
Based on the prediction by PCFNet, we determine the location and membership of the protocluster as follows.
First, we select the protocluster member candidates with a higher probability than a certain threshold, $\sigma_{\mathrm{prob}}=\sigma_\mathrm{{th}}$ ($\sigma_{\mathrm{prob}}$ is defined in the Eq. \ref{eq:sigma} below).
Next, the surface number density map of the selected protocluster member candidates is measured with apertures of $r=1.8'$ at $1'$ intervals across the entire area. 
The center of the protocluster candidates is then determined by applying peak detection to the map.
We use a Python library called \texttt{findpeaks} \citep{Taskesen2020} that utilizes persistent homology for peak detection. 
After the peaks are detected, the protocluster member candidates are assigned to the nearest peaks.
Then, following \citet{Toshikawa2024}, we consider groups within $8'$ of each other as sub-structures.
Note that it is possible that two or more protoclusters overlap on the sky.
Finally, we consider groups of $N_{\mathrm{mem}}=3$ or more member candidates to protoclusters. 
The threshold is the minimum number of protocluster members, i.e., $\gamma N_{\mathrm{th}}$ in case of lowest completeness.

\section{Result}
\label{sec:result}

\subsection{Detection performance per galaxy}
\label{ssec:dppg}
\begin{figure}[tbh!]
    \centering
    \includegraphics[width=0.45\textwidth]{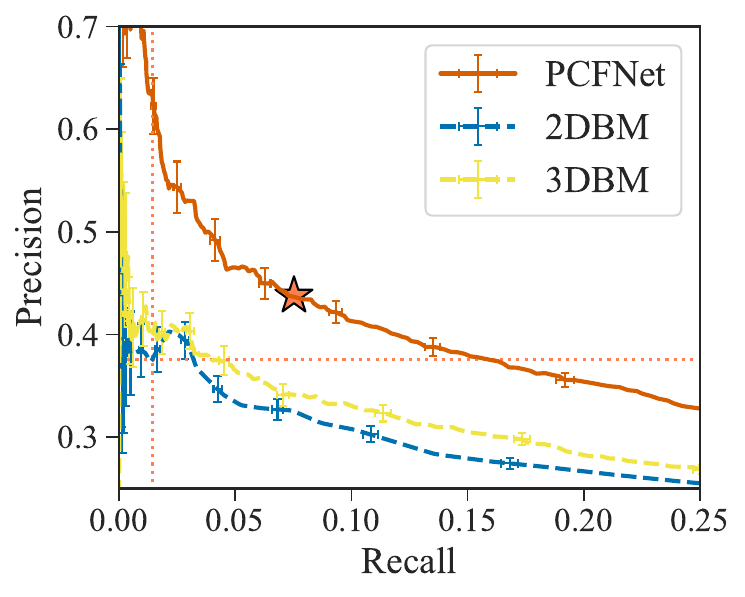}
    \caption{PR curve for protocluster member detection at $z\sim4$. The horizontal and vertical axes represent the recall and precision, respectively. The solid orange line represents the PR curve of PCFNet, the blue line represents that of the surface number density-based model (2DBM), and the yellow line represents that of the integrated probability density-based model (3DBM). The orange dotted line represents the recall and precision at $4\sigma$ for 2DBM. The orange stars with black lines represent the point of PCFNet at $\sigma_{\mathrm{th}} = 2.5$.}
    \label{fig:prcurve}
\end{figure}

Since the number of protocluster member galaxies is smaller than the number of field galaxies that do not belong to protoclusters, the data are imbalanced.
Here we use robust metrics: Precision-Recall curve (PR curve, Figure \ref{fig:prcurve}) and Precision-Recall Area Under of Curve (PR AUC, Table \ref{table:prauc})\footnote{Precision and recall are defined as $\mathrm{TP/(TP+FP)}$, $\mathrm{TP/(TP+FN)}$, where TP, FP, and FN are the number of True Positive, False Positive, and False Negative, respectively. Precision represents how completely selected, and Recall does how purely selected.}. 
We evaluate the errors using the bootstrap method by repeating the resampling $b=100$ times. 
We also evaluate the surface number density-based model (2DBM) and the integrated probability density-based model (3DBM) for comparison. 
2DBM is a similar method of \citet{Toshikawa2018}, which uses a surface number density of the dropout galaxies with a $1.8'$ aperture on the sky to proxy the probability of being a member galaxy. 
3DBM is the simplest method into which the distance in line-of-sight is incorporated. 
We use the PDF estimated by MDN and obtain the maximum number density weighted by the probabilities in line-of-sight.
Compared with the PR AUC, PCFNet is more accurate than the 2DBM and 3DBM by about 0.05 points.

We define the significance of the probability $p_i$ output by PCFNet as follows,
\begin{equation}
    \sigma_{\mathrm{prob}}=\dfrac{p_i-\langle{p}\rangle}{\mathrm{Std}(p)},\label{eq:sigma}
\end{equation}
where $\langle{p}\rangle$ and $\mathrm{Std}(p)$ are mean and standard deviation of evaluation data.
Using $\sigma_{\mathrm{prob}}\geq2.5$ as the threshold, the precision and recall of PCFNet are $7.5\pm0.2\%$ and $44\pm1\%$, respectively.
On the other hand, the precision and recall of 2DBM are $1.5\pm0.1\%$ and $38\pm2\%$ respectively with the threshold as $\sigma_{\mathrm{prob}}=4$, that \citet{Toshikawa2018} used.
The recall of PCFNet with the equivalent precision of 2DBM with $4\sigma$ threshold is $16\pm2\%$, which is approximately 11 times the recall of the 2DBM.

One potential reason for the advantage of PCFNet over 2DBM is that PCFNet can take into account the line-of-sight distance of each dropout galaxy when calculating the protocluster member probability.
The distance is useful to eliminate fore- or background contamination and increase the significance of overdensity.
The brightness of galaxies could be another determining factor for PCFNet in distinguishing protocluster member galaxies since protocluster member galaxies tend to be brighter than field galaxies \citep{Ito2020}. The impact of brightness is discussed in section \ref{ssec:uv}.

\begin{table}
    \centering
    \begin{tabular}{ccc}\hline
     PCFNet& 2DBM & 3DBM\\\hline\hline
     $0.282\pm0.005$& $0.226\pm0.003$& $0.235\pm0.004$\\\hline
    \end{tabular}
    \caption{Comparison of PR AUC for protocluster member detection of PCFNet, 2DBM, 3DBM.}
    \label{table:prauc}
\end{table}

\subsection{Detection performance per protocluster}
\label{ssec:dppp}
\begin{figure}
    \centering
    \includegraphics[width=0.45\textwidth]{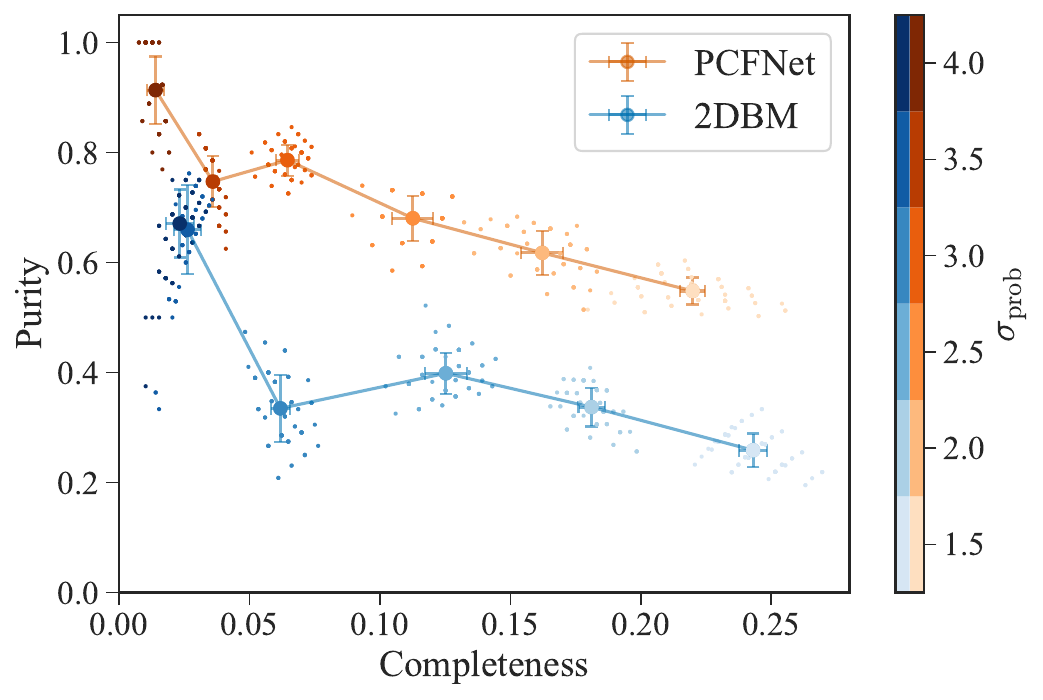}
    \caption{Detection performance per protoclusters at $z\sim4$. The horizontal and vertical axes represent completeness and purity, respectively. The solid lines connecting the dots show the transition of the protocluster detection performance when the threshold for detecting protoclusters is varied as $\sigma_\mathrm{th} = $ 1.5, 2.0, 2.5, 3.0, 3.5, and 4.0.
    The orange and blue respectively represent PCFNet and 2DBM, and the darker color indicates a higher threshold of detecting protoclusters. The smaller dots indicate each of the $b = 100$ bootstrap samples.}
    \label{fig:comppurity}
\end{figure}

We measure the completeness and purity of protocluster detection for our method at $z\sim4$.
The completeness is defined as the number of detected protocluster candidates divided by the number of actual protoclusters in the evaluation data.
The purity is defined as the number of correctly detected protocluster candidates divided by the number of detected protocluster candidates.
We define correctly detected protoclusters as groups whose member fraction belonging to actual protoclusters is greater than 29\%, which is twice the mean fraction of protocluster member galaxies in the simulation data.
We note that even if we change the proportion, the following results do not change significantly.
Figure \ref{fig:comppurity} shows the completeness and purity of protocluster detection when the threshold of protocluster member galaxies is varied as $\sigma_{\mathrm{th}}=1.5,\,2.0,\,2.5,\,3.0,\,3.5,\,4.0$.
The uncertainties are evaluated using the bootstrap method. We repeat the sampling of four light cones from the evaluation data with replacement $b=100$ times.
The points for each bootstrap sample are also shown as small dots in Figure \ref{fig:comppurity}.
From Figure \ref{fig:comppurity}, the purity of 2DBM decreases immediately beyond $\sigma=3.5$, while our model maintains higher purity.
The purity of our model with the threshold as $\sigma_{\mathrm{th}}=2.5$ also exceeds that of 2DBM with the threshold as $\sigma_{\mathrm{th}}=4.0$, which means that it is possible to increase completeness while maintaining purity.
In the following sections, we adopt $\sigma_{\mathrm{th}}=2.5$, which can achieve equivalent purity in the previous work \citep{Toshikawa2018} and high completeness.
At this threshold, the completeness and purity of our model are $10.9\pm0.8\%$ and $69\pm4\%$, respectively\footnote{The completeness and purity of the 2DBM with $\sigma_{\mathrm{th}}=4.0$ are $2.3\pm0.6\%$ and $67\pm9\%$, respectively.
This value is slightly smaller than the result of \citet{Toshikawa2018}, where the purity is $76 \%$, but this might be due to a different definition of protocluster.}.
Note that changing the threshold of the significance of the probability enables us to control the purity and completeness according to purpose.

\section{Application to Observational data}
\label{sec:applyobs}
\subsection{Observational Data -- HSC-SSP}
\label{ssec:obsdata}
We apply PCFNet to the photometric catalog of the HSC-SSP PDR3 S20A Deep/UltraDeep of $37 \mathrm{\,deg^2}$ \citep{Aihara2022}, an optical multi-wavelength large-scale imaging observation program conducted by the Subaru Telescope. HSC-SSP S20A is classified into three regions, Wide, Deep, and UltraDeep, according to the survey depth. The Deep/UltraDeep region is available in multi-wavelength data combined with data from other telescopes and is suitable for validating the deep learning model to detect protoclusters. On the other hand, the Wide region, which has a large survey area ($\sim900\mathrm{\,deg^2}$), is the best region to actually apply the deep learning model to build a statistical sample of protoclusters; for more information on HSC-SSP, see \citet{Aihara2022}.

We use the \texttt{forced} table in the HSC-SSP catalog, which contains the same position photometry in all bands, and adopt the convolved point spread function (PSF) photometry.
Since the photometry of HSC-SSP has systematic offsets from those of the other observation in each band \citep{Aihara2022}, we correct this by using \texttt{\{g,r,i\}\_mag\_offset}, \texttt{gir\_offset} in the HSC-SSP catalog.

In order to prevent false detections, we use \texttt{pixelflags} to remove those affected by \texttt{bad pixel}, \texttt{saturated}, and \texttt{brigthstar} with pixel flags used in \citet{Ono2018}. 
Also, the objects affected by ghost, halo, and blooming that occur around bright objects are removed by the \texttt{mask\_brightstar\_\{ghost15,halo,blooming\}} flags.
In addition, to guarantee that the photometric values are correct, we assign the condition \texttt{\{r,i\}\_blendedness\_abs} $<0.2$ to ensure that the fraction of flux affected by the surrounding objects is less than 20\%.
The data used are restricted to regions where \texttt{\{g,r,i,z,y\}\_inputcount\_value} $N_c$ satisfies $N_c\geq(10,10,10,10,10)$ in all $(g,\,r,\,i,\,z,\,y)$ bands. 
In order to distinguish between Deep and UltraDeep layers, the regions with $N_c\geq(17,16,27,47,62)$ in the COSMOS region and $N_c\geq(13,13,27,42,38)$ in the XMM-LSS region are defined as UltraDeep layer, and the other regions are defined as Deep layer \citep{Ono2018}.
All flags used are summarized in Table \ref{table:obsflag}.
Furthermore, we 
remove areas affected by the stray light or failed aperture correction.
Moreover, we remove the edges of the area, which have a higher flux error than $3\sigma$ to secure uniform depths.
Finally, to detect protocluster members conservatively, we restrict the galaxies bright enough with $(r<m_{\lim,5\sigma})\land (23<i<26)$. The bright-end limit is to avoid contamination from faint stars or low-$z$ objects.
Eventually, we impose the dropout selection (Eq. \ref{eq:gdropout}) on the observational data as simulation data (see Sec \ref{ssec:lbg}) and select 114286, 24970 $g$-dropout galaxies from the HSC-SSP Deep and UltraDeep layers, respectively.

\begin{table*}[tbh!]
\setlength{\tabcolsep}{1mm}
\centering
\begin{tabular}{lccl}
\hline
Parameter Name & Value & Band & Note \\
\hline
\hline
\texttt{isprimary} & True & - & Object has no deblended children  \\ 
\texttt{sdsscentroid\_flag} & False & $r,i$ & General failure flags about sdsscentroid \\ 
\texttt{pixelflags} & False & $g,r,i,z,y$ & General failure flags about pixelflags\\ 
\texttt{merge\_peak} & False & $r,i$ & peak detected in the filters \\ 
\texttt{mask\_brightstar\_ghost15} & False & $g,r,i,z,y$ & Object is within the 1.5 times ghost mask \\ 
\texttt{mask\_brightstar\_halo} & False & $g,r,i,z,y$ & Object is within the halo mask \\ 
\texttt{mask\_brightstar\_blooming} & False & $g,r,i,z,y$ & Object is within the blooming mask \\ 
\texttt{blendedness\_abs} & $<0.2$ & $r,i$ & how much flux is affected by others \\ 
\texttt{inputcount\_value} & $\geq10$ & $g,r,i,z,y$ & the number of images \\
\texttt{apertureflux\_20\_flag} & False & $g,r,i$ & General failure flags about apertureflux \\ 
\texttt{apertureflux\_20\_flaxerr} & NaN & $g,r,i$ &  flux uncertainty\\ 
\texttt{convolvedflux\_0\_20\_flag} & False & $g,r,i$ & General failure flags about convolvedflux \\ 
\texttt{convolvedflux\_0\_20\_flaxerr} & NaN & $g,r,i$ & flux uncertainty\\
\hline
\end{tabular}
\caption{Flags for galaxy selection from the HSC-SSP catalog.}
\label{table:obsflag}
\end{table*}

To identify masked regions around bright stars, we use the HSC-SSP random catalog, which contains uniformly distributed objects with a density of $100\mathrm{\,arcmin^{-2}}$.
To match the real observational data, we impose the flags of \texttt{mask\_brightstar\_\{ghost15,halo,blooming\}}, \texttt{inputcount\_value} to the random catalog and remove the random sources in the area affected by stray light. 
We determine a region to be a masked region if the density of random sources in the regions is less than the average density of  $100\mathrm{\,arcmin^{-2}}$.
We finally use galaxies as targets if less than 50\% of the area within a five arcmin radius centered on a galaxy is in the masked regions: These are 109418 and 23832 in the HSC-SSP Deep and UltraDeep layers, respectively.

\subsection{Apply PCFNet to the HSC-SSP Deep and UltraDeep}
\begin{figure}[!tb]
  \centering 
  \includegraphics[width=0.45\textwidth]{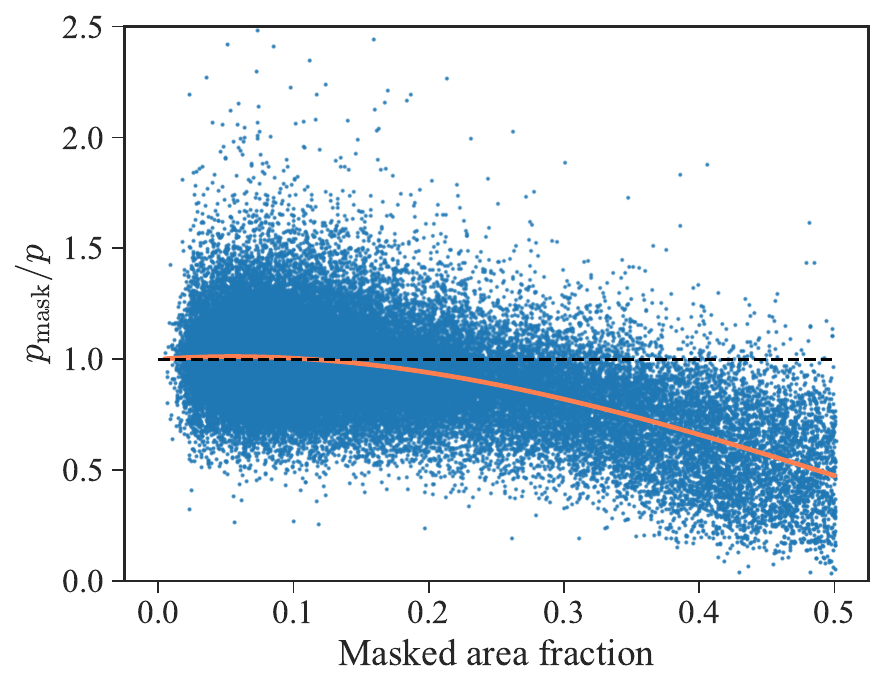}
  \caption{The ratio of the protocluster member probabilities of the mock data with ($p_\mathrm{mask}$) and without ($p$) masked regions. The horizontal axis represents the percentage of the mask within the field of view (within a $5'$ radius), and the vertical axis represents the ratio to the case without any masked regions. The orange line shows the fitting curve as a cubic function.}
  \label{fig:maskaffect}
\end{figure}

We apply PCFNet to the observational data, HSC-SSP Deep, and obtain the protocluster member probabilities for each galaxy.
One thing should be noted before applying it to observed data: unlike simulated data, actual observation data has masked regions.
Masked areas reduce the number of galaxies, which in turn reduces the performance of PCFNet and decreases the protocluster member probability.
To quantify the effect, we first make the mock mask regions on the simulation data. 
The mask regions of the COSMOS UltraDeep region, which has the same field of view as the PCcone light cone, are directly converted to mock mask regions on PCcone.
We exclude PCcone galaxies that are within $10^{-4}$ arcmin of the nearest random object in the mock mask region.
For each target galaxy, the masked area fraction is calculated from the number of random objects within a five arcmin radius of the galaxy, and target galaxies with a fraction of 50\% or less are excluded.
We then compute the ratio of the protocluster member probabilities between the mock data with and without mask regions as a function of the masked area fraction (Figure \ref{fig:maskaffect}).
As expected, the probability decreases as the mask fraction increases, although the variance is large.
We approximate this function with a cubic function to adjust the protocluster member probability.
Protoclusters whose masked area fraction exceeds 30\% are excluded because the correction is too large.

\begin{figure*}[tbh!]
    \centering
    \includegraphics[width=0.95\textwidth]{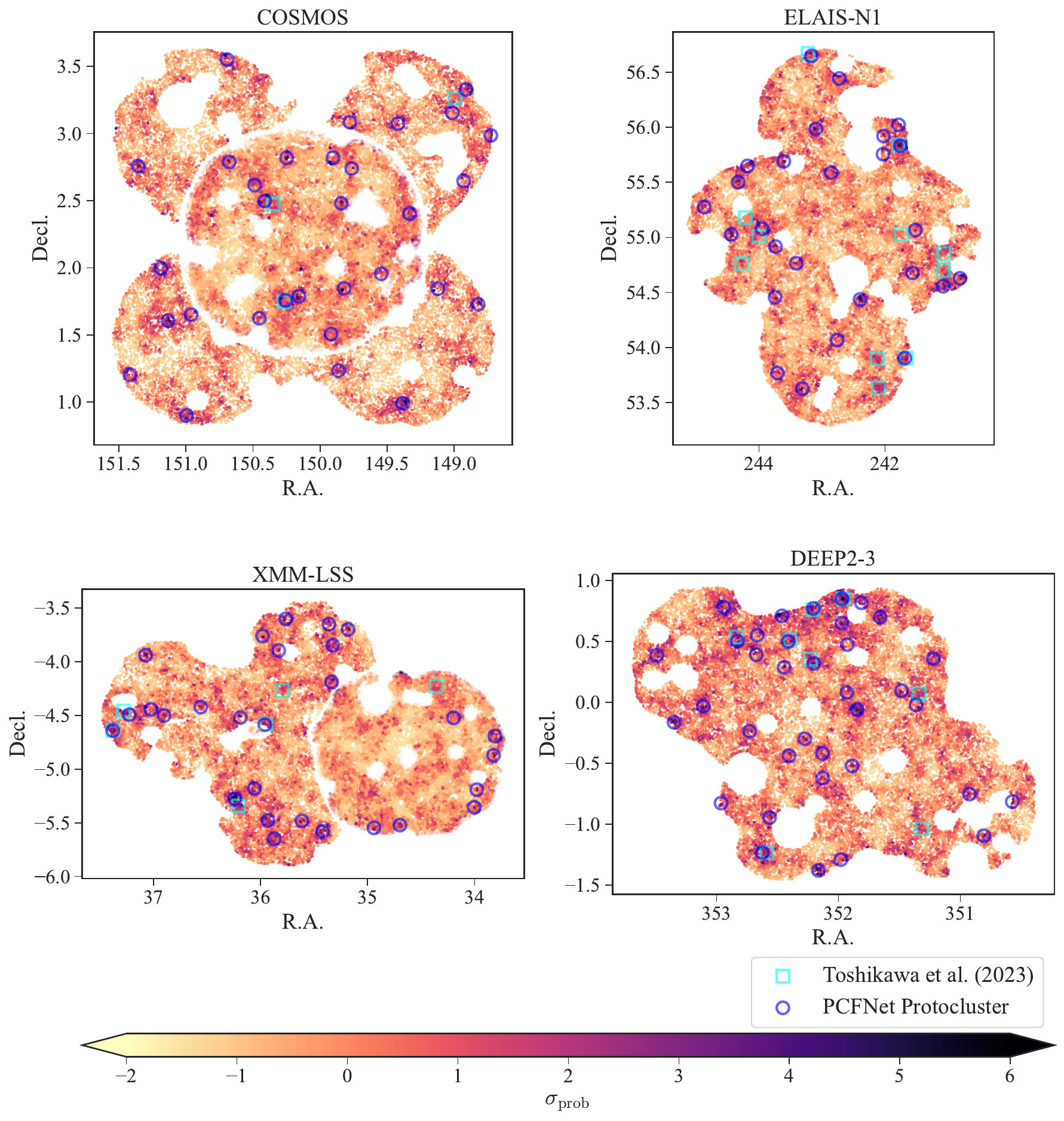}
    \caption{Sky distribution of the predicted probabilities of protocluster member galaxies and protocluster candidates at $z\sim4$ in the Deep and UltraDeep layers. Each point represents a galaxy, and the color indicates the probability of being a protocluster member galaxy predicted by PCFNet. 
    The blue circles and cyan squares indicate the positions of the detected protocluster candidates by PCFNet and \citet{Toshikawa2024}, respectively.}
    \label{fig:obsdist_all}
\end{figure*}

\begin{figure*}[tbh!]
    \centering
    \includegraphics[width=0.45\textwidth]{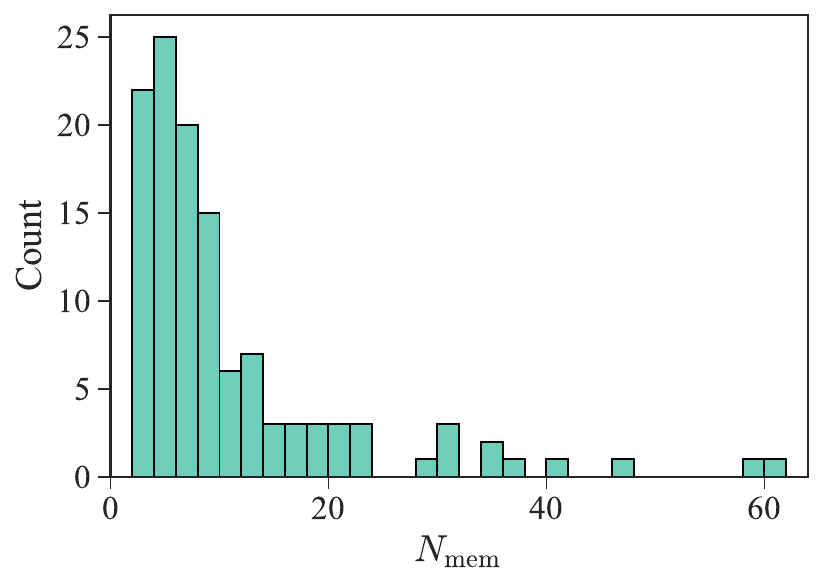}
    \includegraphics[width=0.45\textwidth]{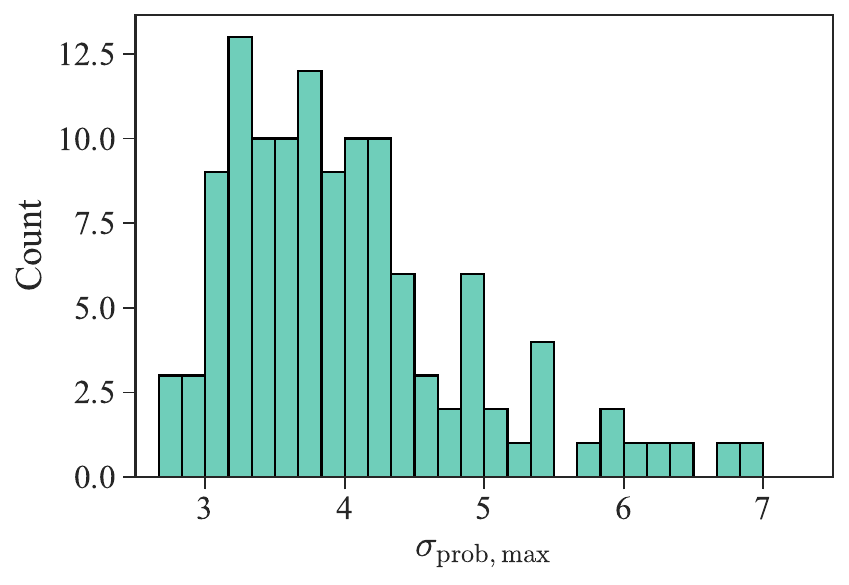}
    \caption{The distributions of the number (left) and the maximal significance (right) of members of protocluster candidates at $z\sim4$ in HSC-SSP Deep and UltraDeep layers.}
    \label{fig:dist_n_sigma_pcs}
\end{figure*}

The significance, $\sigma_{\mathrm{prob}}$, is derived from the protocluster member probabilities using Eq. \ref{eq:sigma} with $\mean{p},\,\mathrm{Std}(p)$, which are obtained from the evaluation data. Then we identify the protocluster candidates by applying $\sigma_{\mathrm{th}}=2.5$. 
We find 121 protocluster candidates from $\sim17.6\,\mathrm{deg}^2$ of the HSC-SSP Deep and UltraDeep layers.
Figure \ref{fig:obsdist_all} shows distributions of galaxies and protocluster candidates. 
Note that we independently find protoclusters from the Deep and UltraDeep layers to avoid domain shift since they have different number densities and depths.
The effective area of the target region and the number of protocluster candidates are summarized in Table \ref{table:effective_area}.
Figure \ref{fig:dist_n_sigma_pcs} shows the distribution of the number and the maximum significance of the protocluster member probabilities of the candidate members.
The protoclusters that have the member with high probability ($3.0\mathchar`- 4.5\sigma$) are the majority (74\%). 
We also note that the detected protoclusters are only candidates and have not yet been confirmed. 
In the following, we refer to the protocluster candidates found by PCFNet as detected, regardless of whether they are observed or simulated.
On the other hand, an actual protocluster means that its three-dimensional structure is identified, and it will evolve into a cluster by $z=0$; it is mainly used in the simulation.

\begin{table*}[tbh!]
  \centering
  \begin{tabular}{ccccc}
    \hline
    &region & area [deg$^2$] & $N_{\mathrm{pc}}$ & number density [deg$^{-2}$]\\
    \hline
    \hline
    Deep & COSMOS & 2.99 & 17 & 5.7\\
     & DEEP2-3 & 4.37  & 36 & 8.2\\
     & ELAIS-N1 & 4.10  & 26 & 6.3\\
     & XMM-LSS & 2.83 & 21 & 7.4\\
     \hline
    UltraDeep & COSMOS & 1.69  & 14 & 8.3\\
     & XMM-LSS & 1.61  & 7 & 4.3\\
    \hline
  \end{tabular}
  \caption{Summary of the effective area, the number, and number density of detected protoclusters at $z\sim4$ in each field.}
  \label{table:effective_area}
\end{table*}

The number of detected protoclusters per effective area is $6.9\,\mathrm{deg^{-2}}$.
This is 3.8 times larger than the number of detections in \citet{Toshikawa2024}, who used a method similar to the 2DBM.
This is consistent with the predicted yield of $4.7\pm1.3$ compared between 2DBM and PCFNet in PCcone within the error (see \ref{ssec:dppp}).
Although it is difficult to compare different methods for selecting protocluster candidates in a situation where the true protocluster is not known, our 63\% protocluster candidates match the protoclusters detected by \citet{Toshikawa2024} within $8'$ in the sky.
The other protocluster candidates are not detected by PCFNet here because the protocluster member probabilities are relatively low and do not exceed the detection threshold. 
In fact, by reducing the threshold to $\sigma_{\mathrm{th}}=1.8$, all of the protocluster candidates detected by \citet{Toshikawa2024} are also detected by PCFNet.
On the contrary, those found by PCFNet but not found in \citet{Toshikawa2024} with $4\sigma$ surface overdensity can be low-mass protoclusters ($M_\mathrm{halo}^{z=0}\sim10^{14}M_\odot$), as described in sec. \ref{ssec:mass}.

\section{Discussion}\label{sec:discussion}
\subsection{Descendant Halo mass at $z\sim0$}
\label{ssec:mass}
\begin{figure}[tbh!]
    \centering
    \includegraphics[width=0.45\textwidth]{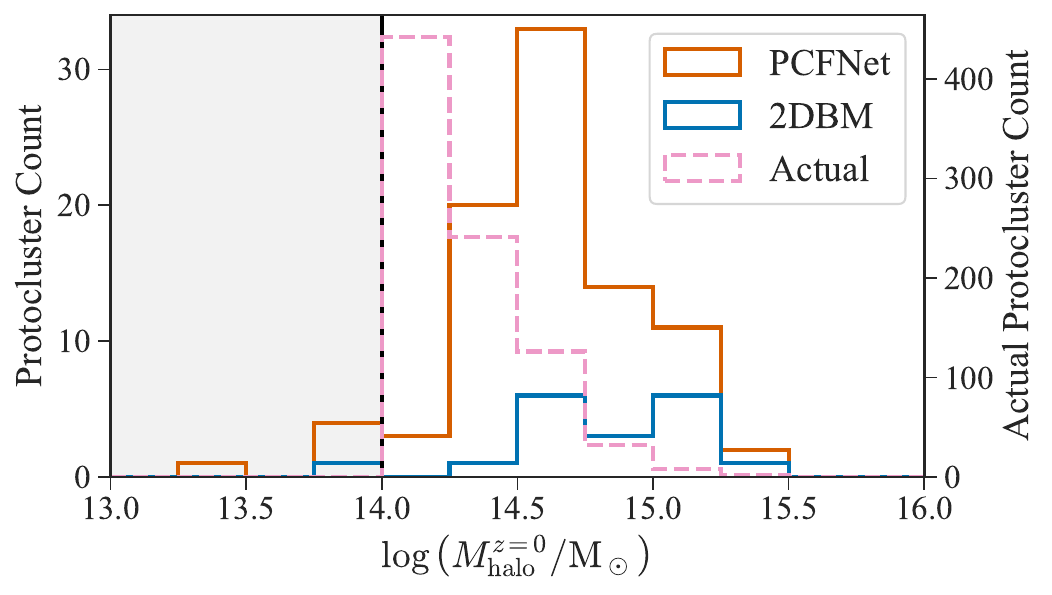}
    \caption{Distribution of halo mass of protoclusters at $z=0$ in the simulation data. The solid lines represent the distribution of maximum halo masses at $z = 0$ of the protocluster candidates detected by each model, with orange and blue histograms representing PCFNet and 2DBM, respectively. The dashed pink line represents the distribution of maximum halo masses at $z=0$ of actual protoclusters in the evaluation data. The gray region indicates that the halo mass at $z=0$ is lower than the protocluster threshold.
    }
    \label{fig:halomass}
\end{figure}

We investigate the descendant halo mass of protocluster candidates and compare them with that of 2DBM in simulation data.
We determine the maximum halo mass of member galaxy at $z=0$ for each protocluster candidate in simulation data and consider it as the represent descendant halo mass of the protocluster candidate at $z=0$.
Figure \ref{fig:halomass} compares the descendant halo mass distribution detected by PCFNet and 2DBM.
PCFNet detects $7\pm2$ times more protoclusters whose halo mass are lower than $M_\mathrm{halo}^{z=0}=10^{15}M_\odot h^{-1}$ than 2DBM.
Moreover, while 2DBM detects only one low-mass protocluster ($10^{14}M_\odot h^{-1}\leq M_{\mathrm{halo}}^{z=0}\leq 10^{14.5}M_\odot h^{-1}$) from four light cones for evaluation, PCFNet achieves to detect 23.

This result indicates that PCFNet, which increases the number of detections while minimizing false detection of protoclusters as shown in Sec \ref{ssec:dppg}, is able to detect even relatively low-mass protoclusters.
We conclude that PCFNet can detect more general protoclusters because the mass distribution of protoclusters is sufficiently extended toward the low-mass side (see the "Actual" curve in Figure \ref{fig:halomass}).

\subsection{The properties of protocluster members}

\subsubsection{rest-UV magnitude on simulations and observations}
\label{ssec:uv}
\begin{figure*}[tbh!]
    \centering
    \includegraphics[width=0.45\textwidth]{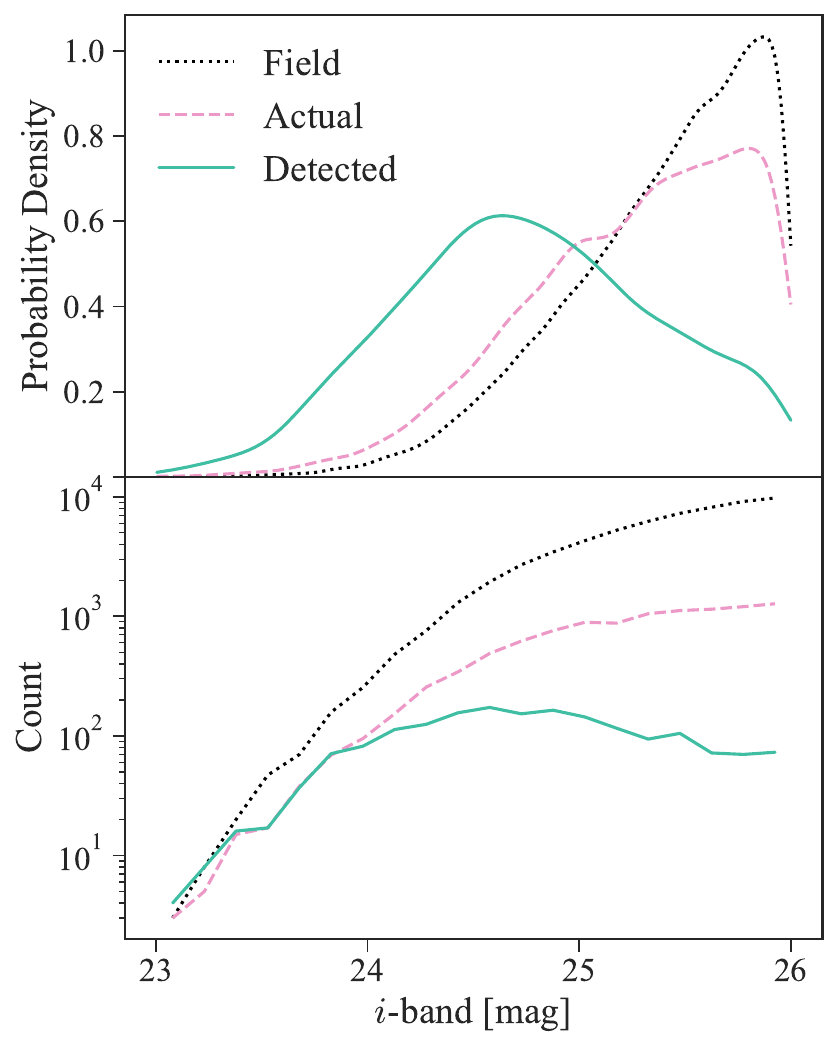}
    \includegraphics[width=0.45\textwidth]{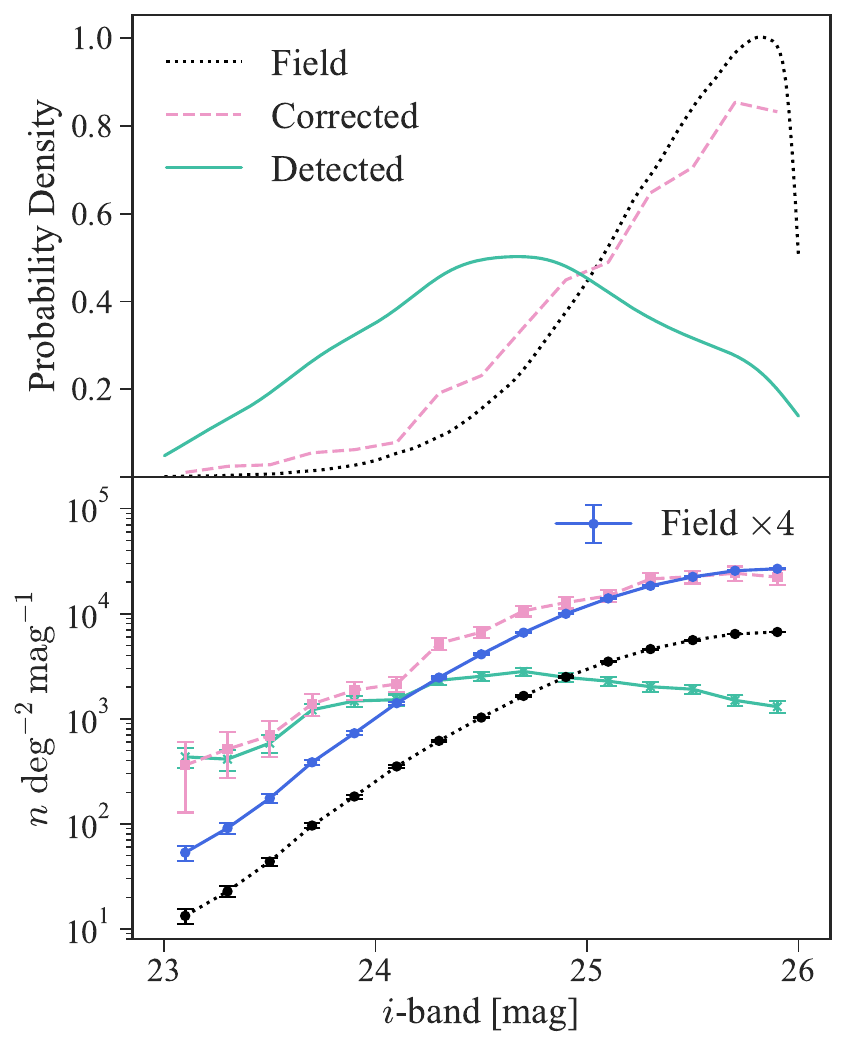}
    \caption{(left) Distribution of $i$-band magnitude of protocluster member galaxies that were detected by PCFNet (green; Detected), non-member field galaxies (blue; Field), and actual protocluster member galaxies (pink; Actual) at $z\sim4$ in PCcone. 
    (right) Similar to the left, but in observational data. The blue line represents the distribution of field galaxies not detected as protocluster member galaxies (Field), and the green line represents the distribution of protocluster member galaxies detected by PCFNet (Detected). The pink line shows the distribution corrected for model bias using Eq. \ref{eq:corr} (Corrected) and the distribution of field galaxies (Field $\times 4$) multiplied by 4.0 to match the scale on the faint side.
    The error bars of each line indicate Poisson errors.
    The curves in both upper figures are probability density functions for each normalized distribution.}
    \label{fig:properties}
\end{figure*}

\begin{figure*}[tbh!]
    \centering
    \includegraphics[width=0.45\textwidth]{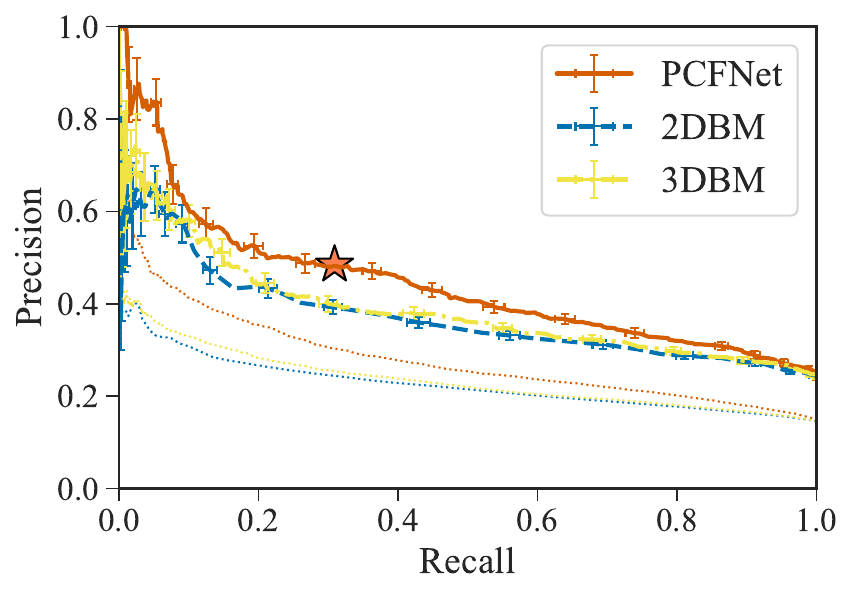}
    \includegraphics[width=0.45\textwidth]{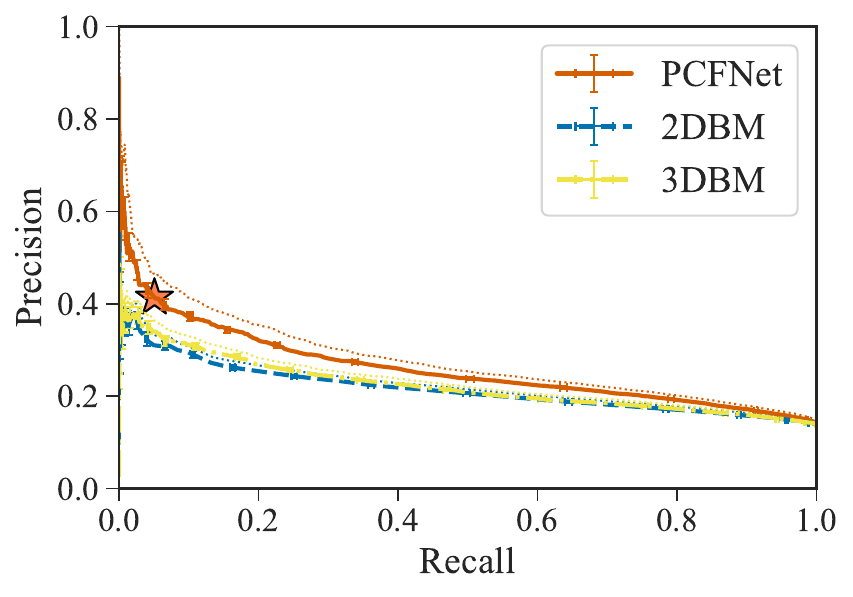}
    \caption{Comparison of PR curves between different brightness (left: $i \leq 24.5$, right: $i > 24.5$) in PCcone.
    The orange, blue, and yellow lines represent the PR curve of PCFNet, the surface number density-based model (2DBM), and the integrated probability density-based model (3DBM), respectively. 
    The PR curves for all galaxies in Figure \ref{fig:prcurve} are also shown as dotted lines. Note that the $2.5\sigma$ threshold indicated by the orange star with black lines is calculated from all galaxies.}
    \label{fig:comp_prcurve}
\end{figure*}

We examine the properties of the rest-UV magnitude of protocluster members detected by PCFNet from both the simulation and observational data.
Figure \ref{fig:properties} shows the distribution of rest-UV $i$-band magnitude for detected protocluster members, actual protocluster members, and field galaxies in the simulation data.
In the distribution of rest-UV magnitude, an overabundance trend is seen in the detected protocluster members.
A possible reason for this is a bias due to the brightness of the detected protocluster members.
The number of brighter ($i<24$) members in the detected protoclusters is consistent with that of the actual protocluster members.
This means PCFNet detects the almost bright galaxies in protoclusters.
On the other hand, the faint galaxies are less detected by PCFNet.
To see this trend more clearly, we compare two PR curves of the bright ($i<24.5$) and faint ($i>24.5$) galaxies in Figure \ref{fig:comp_prcurve}. 
It is apparent that brighter galaxies have better detection performance, with a roughly six-fold difference in the recall with a threshold of $\sigma_\mathrm{th} = 2.5$ calculated for the entire galaxies.
Thus, it should be noted that a selection bias arises in the rest-UV magnitude distribution by using the same threshold for targets with different luminosity.

While we use simulated data in the above discussion, we review trends in observational data.
In order to make comparisons of the surface number density, we use the members of the detected protocluster candidates only. 
In other words, isolated protocluster members are not included in the analysis.
The area used to determine surface density is the entire study area for field galaxies and the total protocluster area for protoclusters.
A single protocluster area on the sky is calculated from the circle defined by the following radius, $r_\mathrm{pc}$.
\begin{equation}
    r_\mathrm{pc} = \dfrac{1}{N_\mathrm{mem}}\sum_i \left|\Vec{x}_i - \Vec{x}_G\right| \label{eq:radius},
\end{equation}
where $\Vec{x}_G$ represents the center of the protocluster.

The right panel of Figure \ref{fig:properties} is similar to the left one but shows the $i$-band magnitude distribution for observational data.
The number density of detected protocluster members decreases at the faint end.
This is attributed to the following two entangled factors: The first is that there are more bright galaxies in the protoclusters than in the field as the intrinsic environmental effect, and the second is that the faint galaxies are relatively hard to detect with PCFNet as an artificial effect (Figure \ref{fig:properties}, left), so-called selection bias.

We correct the selection bias of PCFNet by using simulation data.
First, we compute the $i$-band magnitude distribution ratios between the detected $(\phi^{\mathrm{d}}_\mathrm{pc,s})$ and actual $(\phi^{\mathrm{t}}_\mathrm{pc,s})$ protocluster member galaxies in simulation data. 
These ratios represent the first-order selection bias. 
Therefore, the number density function by $i$-band magnitude without the selection bias ($\phi^\mathrm{t}_\mathrm{pc}$) approximate to the raw number density function $(\phi^\mathrm{d}_\mathrm{pc})$ multiplied by the correction factor,
\begin{equation}
    \phi^\mathrm{t}_\mathrm{pc}\approx \hat{\phi}^\mathrm{t}_\mathrm{pc} = \phi^\mathrm{d}_\mathrm{pc} \left(\dfrac{\phi^\mathrm{t}_\mathrm{pc,s}}{\phi^\mathrm{d}_\mathrm{pc,s}}\right). \label{eq:corr}
\end{equation}

Note that this correction does not offset all model-induced biases.
Other possible but inevitable biases are caused by the difference between the simulational and actual universe and by the approximation performance of the neural network.
When considering the protocluster candidates in observational data, it is always important to be aware of the artificial effects caused by the model and differences from the simulation data.

The number density distribution of detected protocluster members corrected with Eq. \ref{eq:corr} is shown in the right panel of Figure \ref{fig:properties} as Corrected, which indicates that the density distribution has an excess at the bright end compared to that of the field galaxies, which is normalized at the faint end. 
This suggests that there are more bright galaxies in the protocluster than in the field at $z\sim4$, consistent in the literature \citep{Ito2020, Toshikawa2024}.
It is inferred from the result that the galaxies in protoclusters are massive and have high SFR at $z\sim4$, and star formation in the protocluster regions is already highly active beyond $z>4$. 
Moreover, considering the discussion in Sec. \ref{ssec:mass}, it is suggested that the environmental effects in protocluster at high redshift are also seen on the lower-mass side.
However, as mentioned above, there is a fair chance for lurking significant potential biases affecting the result, so it is necessary to investigate the influence in more detail.

\subsubsection{stellar mass, SFR, and sSFR on simualtions}

\begin{figure*}[tbh!]
    \centering
    \includegraphics[height=9.5cm]{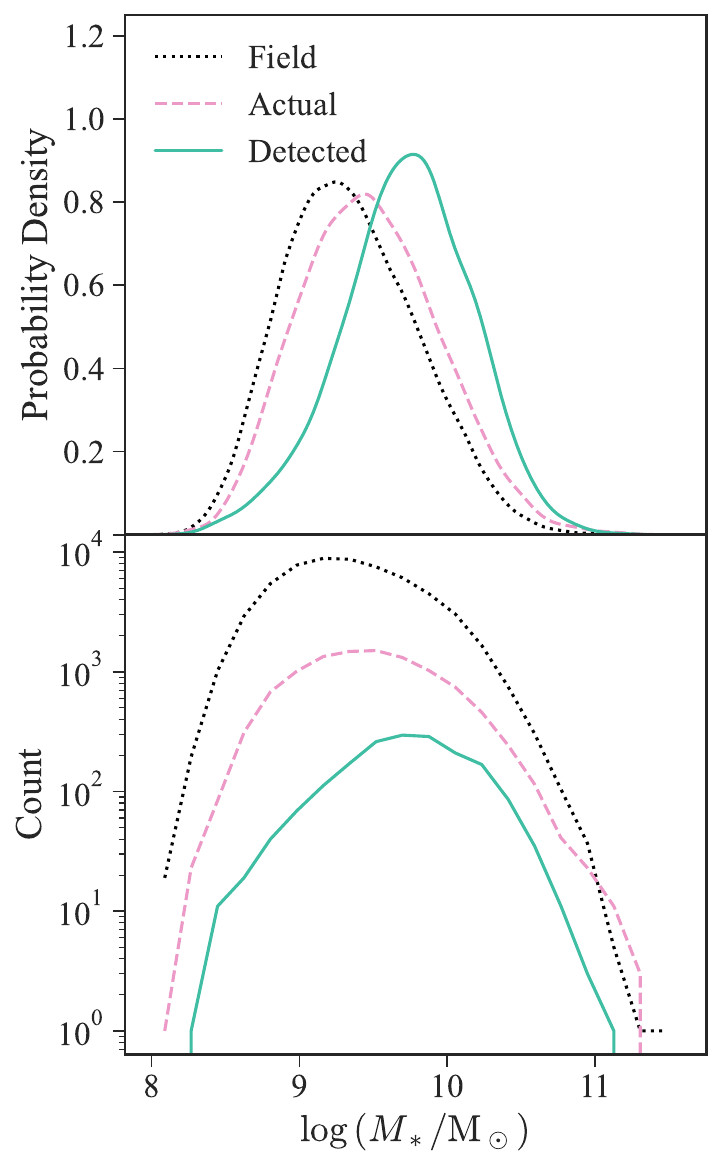}
    \includegraphics[height=9.5cm]{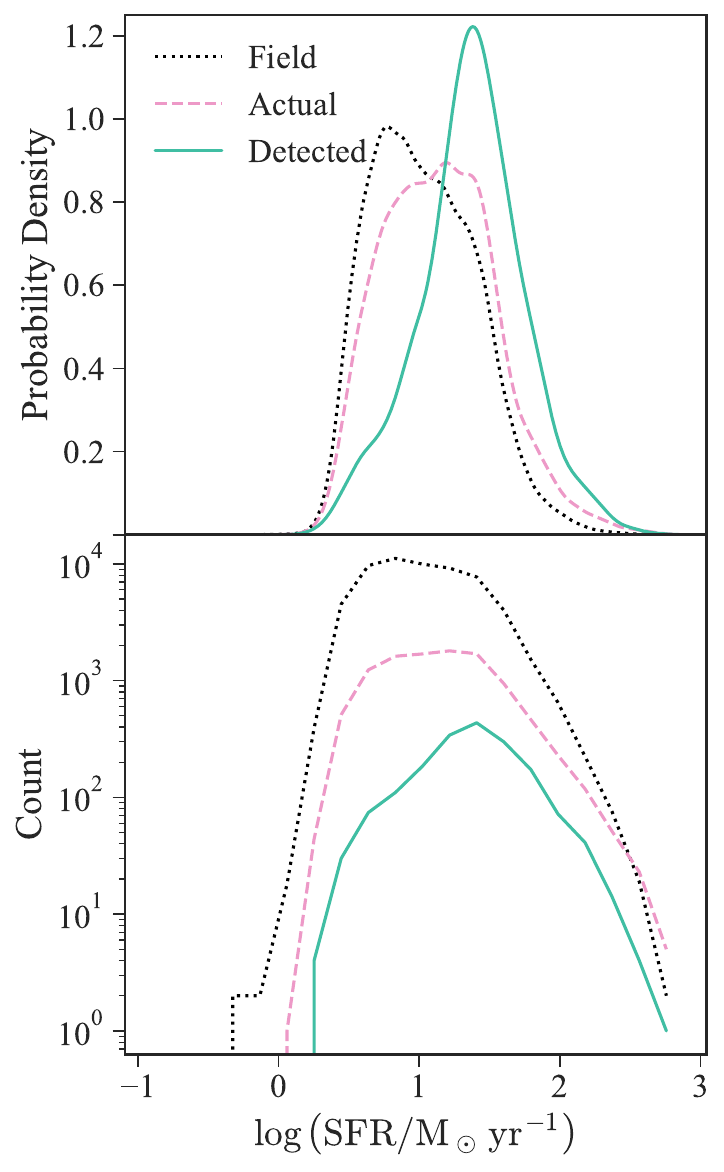}
    \includegraphics[height=9.5cm]{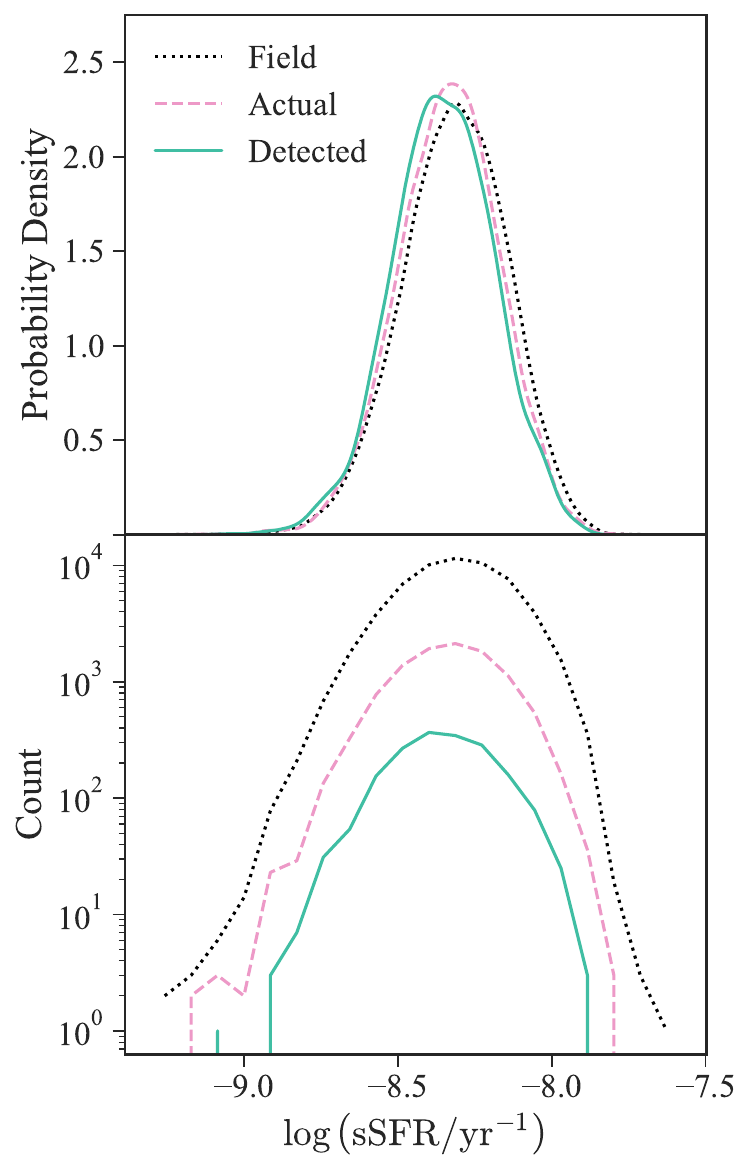}
    \caption{Distribution of the stellar mass, SFR, and sSFR of protocluster members at $z\sim4$ in the simulation data. The legend of colors is the same as the left panel of Figure \ref{fig:properties}.}
    \label{fig:properties_masssfr}
\end{figure*}

While it is hard to obtain the properties of galaxies in observational data which are observed with only a few photometric bands, they are accessible in simulation data.
Figure \ref{fig:properties_masssfr} shows the distributions of stellar mass, SFR, and specific SFR (sSFR) of protocluster member candidates detected by PCFNet, actual protocluster members, and field galaxies at $z\sim4$ in the simulation data.
Note that the contaminated protocluster member candidates from low-$z$ are removed.
Compared to field galaxies, the actual protocluster members show number excesses where both stellar mass and SFR are high.
This is expected given that the rest-UV luminosity is correlated with each of the two.
On the other hand, the sSFR does not significantly differ, and both do not deviate from the main sequence.
Thus, the bright-end excess is not due to starburst galaxies but is presumably due to the simple over-abundance of massive star-forming galaxies.
This result is consistent with \citet{Toshikawa2024}.
These imply the galaxies in protoclusters experience early star formation, which agrees with the pictures predicted by the previous theoretical studies \citep{Muldrew2015, Chiang2017}.
\citet{Staab2024} also shows the clear correlation between SFR and overdensity with the observation of Taralay protocluster at $z\sim4.57$ and a consistent view of stellar mass growth in the early universe with our result.
In the future, the follow-up observation of protocluster candidates in various epochs will provide a more detailed study on protocluster properties binning by total stellar mass, halo mass, and so on.

\subsection{Protoclusters with quiescent galaxies}

Quiescent galaxies (QGs), which are galaxies that have quenched their star formation, are numerous in the nearby universe, mainly in clusters \citep{Dressler1980}.
On the other hand, the census of QGs at $z>2$ is still controversial, and understanding when and how star formation is quenched is one of the main motivations for protocluster studies.
\citet{Muldrew2018} studied the distribution and fraction of QGs in protoclusters using the semi-analytical model of \citet{Henriques2015}, which PCcone also uses as its basis, and reported the presence of QGs inside and outside the main halo of protoclusters even at $z\sim4$.
In this section, we investigate the relationship between the QGs and the core galaxies of the protoclusters defined in this study based on the simulation data. 

Since QGs are not selected by dropout techniques, we have to construct the sample in another way.
First, we select all galaxies with $\mathrm{sSFR<10^{-9.5}}$ as QGs at $3.5<z<4.5$ in PCcone following \citet{Ito2021}. 
Note that the results in this section do not change with the modification to $\mathrm{sSFR} \leq 10^{-9.635}$, which is defined in  \citet{Muldrew2018}.
The 2392 QGs are selected from the 20 light cones. Note that due to PCcone constraints, the QGs have a limiting mass of $M_* \geq 10^8 M_\sun$.
For each QG, the distance to the core galaxy (defined in section \ref{ssec:pc_define}) is calculated at $z\sim4$, and the protocluster member QGs are identified as those with distances to their nearest core galaxies within $r = 5.5 \mathrm{\,cMpc}$.
The number of core galaxies associated with QGs is 319 (1.8\% of all core galaxies).
Note that some core galaxies are associated with more than one QG.

\begin{figure*}[tbh!]
    \centering
    \includegraphics[width=0.45\textwidth]{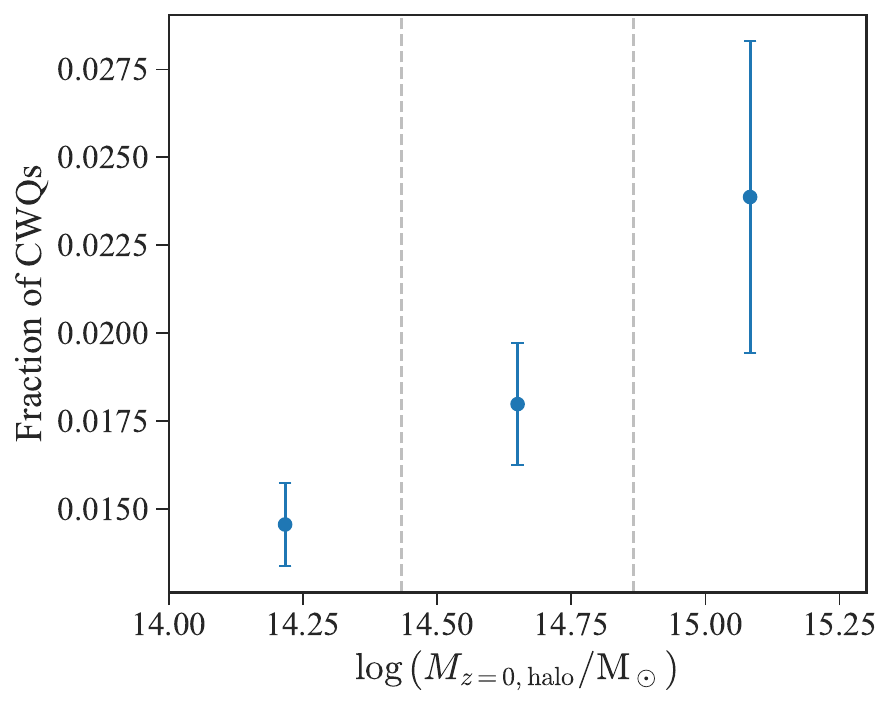}
    \includegraphics[width=0.45\textwidth]{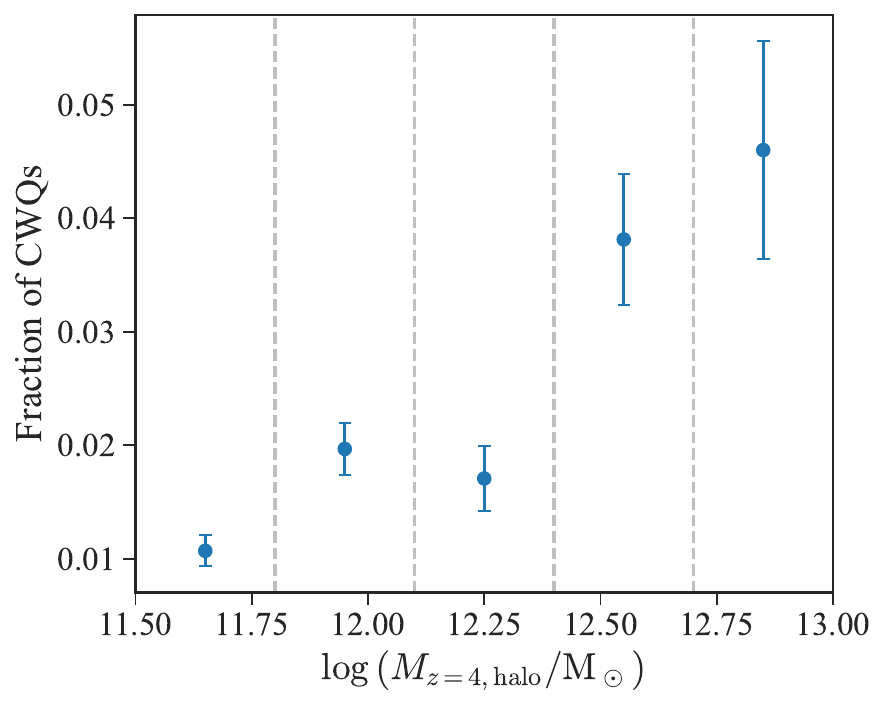}
    \caption{CWQ fraction as functions of halo masses at $z = 0, 4$ in the simulation data. The error bars indicate Poisson error in each bin.
    Note that the fraction of core galaxies associated with QGs relative to the halo mass at $z = 0$ inclines more gradually with error than the others.
    For this reason, the number of bins is reduced to three in the figure.
    }
    \label{fig:corewqg_prop}
\end{figure*}

Figure \ref{fig:corewqg_prop} shows the fraction of core galaxies associated with QGs (hereafter CWQ fraction) as functions of halo masses at $z = 0 \text{ and } z=4$.
CWQ fraction is derived from dividing the number of CWQs by the total number of core galaxies at the given mass of the halo.
The CWQ fraction is found to increase monotonically with both halo masses.

\begin{figure}[tbh!]
    \centering
    \includegraphics[width=0.45\textwidth]{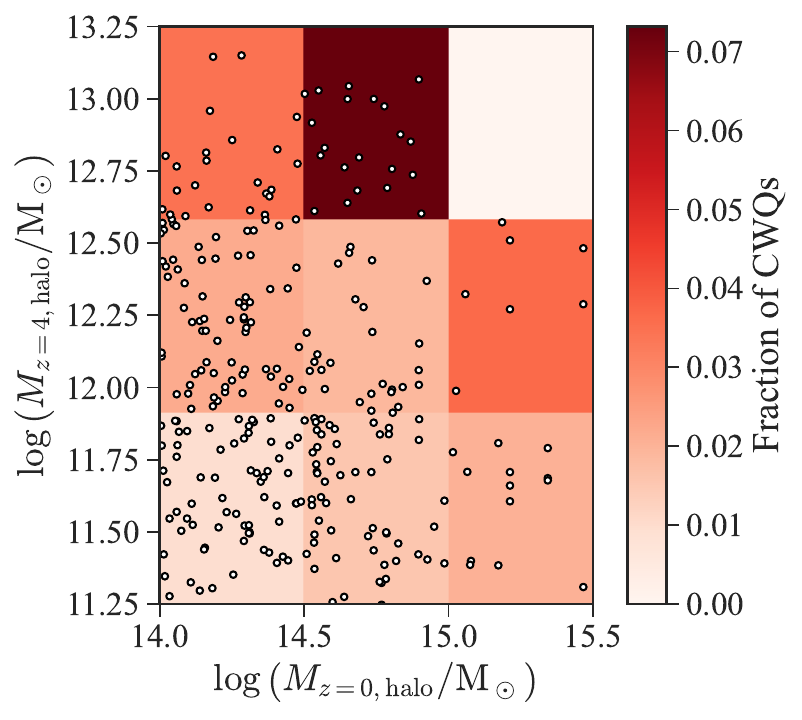}
    \caption{Two-dimensional distribution of halo masses at $z = 0$ and $z = 4$ of CWQs in the simulation data. The color of the background represents the CWQ fraction. The white dots represent the CWQs.}
    \label{fig:corewqg_massz0vs4}
\end{figure}

Figure \ref{fig:corewqg_massz0vs4} shows the relationship between the halo mass at $z = 0$ and $z = 4$ for the CWQs. 
For example, fixing the halo mass bin at $z=0$, the larger the halo mass at $z=4$, the larger the CWQ fraction. 
On the other hand, if we fix the halo mass bin at $z=4$, the larger the halo mass at $z=0$, the larger the CWQ fraction. 
This indicates that the CWQ fraction at $z=4$ is higher not only for halo mass at $z\sim4$, but also for systems that will become more massive in the future.

Given that the high CWQ fraction is a proxy of the rapidity of early galaxy formation, it may suggest that galaxy formation is more accelerated not only in systems with high core galaxy halo mass at $z\sim4$ but also in rich environments around the core galaxy where many halos will be accreting in the future.
This interpretation is consistent with the hierarchical structure formation model \citep{Lacey1993}. 
Another possible interpretation is that the higher the density of the surrounding environment, the earlier the time of quenching of star formation due to environmental effects.

It should be noted that this result is in contrast to \citet{Remus2023}, which studied the evolution of 42 protoclusters at $z\sim4$ in hydrodynamical cosmological simulation suite \textit{Magneticum}\footnote{\url{http://www.magneticum.org}} and found QG fraction does not depend on the virial mass, the member of member galaxies, or the SFR. 
One of the possible reasons for this disagreement is the mass range of the protoclusters under consideration is different: They are targeting the core of protoclusters that are already virialized, and the virial mass has reached $M_\mathrm{vir}>10^{13}M_\odot$ at $z=4.2$.
Due to this difference in definition, our dynamic range for the halo mass is wider.
Another possible reason may be due to the difference between hydrodynamic simulation and semi-analytic model.

Although it is difficult to compare the observed results of QGs at $z\sim4$, which has few observationally detected \citep{Marsan2022, Valentino2023, Carnall2023}, it is possible to approximate the order of the fraction from a certain case.
\citet{Tanaka2023} found a massive QG protocluster which consists of five QGs from Subaru/XMM-Newton Deep Survey \citep[SXDS;][]{Furusawa2008}.
They used the multi-band photometric catalog \citep{Kubo2018} over $0.7\mathrm{\,deg^2}$ and selected QGs from $3.7<z<4.3$.
The halo mass of the most massive QG in the protocluster is estimated as $3.1^{+6.6}_{-2.0}\times10^{12} \,M_\odot$.
The number of core galaxies in the redshift range is estimated $\sim90$ in $0.7 \mathrm{\,deg^2}$ from PCcone, and if restricted core galaxies whose halo masses at $z\sim4$ are more massive than $2\times10^{12}\,M_\odot \, h^{-1}$, the number of core galaxies is $\sim8$, viz. the CWQ fraction is approximated as $\sim0.13$ (0.01  without restriction in mass).
This is consistent or slightly higher with our results ($0.02$) estimated from PCcone, but it should be noted that the estimation has large uncertainties, e.g., only one protocluster is used, the completeness of the method of protocluster detection is not corrected, the protocluster found by \citet{Tanaka2023} only consists of QGs so it is slightly different from protocluster associated with QGs, and so on.
Additional observation in the future would give better insights into the quenching in protoclusters.

Although there is very little that can be said in comparison to observations at present, this study suggests that the wide dynamic range of the mass of protoclusters will be important in future observational studies of the ages or quenching mechanisms in protoclusters.
The detection bias toward massive protoclusters due to the previous method has a detrimental effect on the evaluation of galaxy properties and environmental effects.
As described in section \ref{ssec:mass}, PCFNet is capable of detecting protoclusters of relatively smaller masses than previous studies and is expected to reduce the effect of the bias.

\section{Conclusion}
\label{sec:conclusion}
We develop a new protocluster detection method by deep learning and detect 121 protocluster candidates at $z\sim4$ from HSC-SSP Deep/UltraDeep layer.
We also discuss the properties of detected protocluster from both simulational and observational views. The main contributions of this paper are the following:
\begin{enumerate}
    \item We develop a new deep learning model, PCFNet which is based on PointNet \citep{Qi2016} to handle pointcloud such as galaxy distribution. 
    PCFNet uses relative positions, magnitude, and colors of galaxies within $5'$ around a target galaxy to estimate the probability that the target belongs to a protocluster. 
    We test PCFNet with the semi-analytic model, PCcone \citep{Araya-Araya2021} to achieve the precision and recall are $7.5\pm0.2\%$ and $44\pm1\%$, respectively.
    \item Based on the PCFNet output for each galaxy, protocluster candidates are detected by peak detection from the distribution of protocluster member galaxy candidates with significance greater than $\sigma_\mathrm{th} = 2.5$. 
    The completeness and purity are $10.9\pm0.8\%$ and $69\pm4\%$, respectively.
    \item We apply PCFNet to the HSC-SSP S20A Deep and UltraDeep layers and detect 121 unique protoclusters at $z\sim4$ from $\sim17.6\mathrm{\,deg^2}$.
    \item Comparing the halo mass distribution of protocluster candidates in the simulation data, PCFNet detects $7\pm2$ times more protoclusters with $M_\mathrm{halo}^{z=0}>10^{15}M_\odot h^{-1}$ than previous works which are affected by projection effects and especially 23 lower-mass protoclusters ($M_{\mathrm{halo}}^{z=0} = 10^{14.0-14.5}M_\odot$) among which the previous method detects only one in evaluation data.
    \item We find that the distribution of detected protocluster members is skewed towards brighter and higher in terms of rest-UV luminosity, SFR, and stellar mass.
    The distribution of the rest-UV luminosity of the detected protocluster candidates in observational data also has a bright-end excess, which is consistent with previous observational studies. 
    This suggests that star formation is already active in protoclusters at $z > 4$.
    \item We find the fraction of the core galaxies associated with QGs increases in dependence on halo mass at both $z=0$ and $z=4$ from simulation data.
\end{enumerate}

The protocluster candidates detected in this study are expected to be used for the assessment of environmental effects in protoclusters at high redshift with spectroscopic follow-up observation conducted in the future.
Furthermore, they can be analyzed for each physical property by SED fitting by matching multi-wavelength data, as in \citet{Toshikawa2024}, which will derive detailed discussions with star formation history.

PCFNet can be applied to the HSC-SSP Wide layer, LSST, and other extraordinarily wider surveys with the next-generation telescopes such as Euclid, TMT, Roman, and more protoclusters are expected to be detected in the future.
Moreover, if a new and improved lightcone model based on detailed hydro-dynamical simulation and photometric predictions that are even closer to reality becomes available in the future, we can expect further improvements in PCFNet with updated training data.
In terms of computational resources, a single GPU is sufficient to execute the inference of PCFNet over such a vast survey area in a realistic amount of time (several minutes to several hours), and further acceleration is also easily achieved through parallelization.
Furthermore, the versatility of PCFNet also makes it applicable to other dropout galaxies, and it enables us to search for protoclusters over a wide range of redshifts from $z = 2\mathchar`-8$.

\section*{Acknowledgments}
We appreciate the anonymous reviewer for constructive comments and suggestions that improved the manuscript.
We thank Kazuhiro Shimasaku, Makoto Ando, Tadayuki Kodama, Akito Kusaka, Masato Onodera, Michael Rauch, Shigeki Inoue, Brian Lemaux, Satoshi Kikuta, Hiroki Hoshi, and Shunta Simizu for their valuable comments and Pablo Araya-Araya, Masayuki Tanaka, and the HSC helpdesk for helping us with our analysis.
Y.T. was supported by Forefront Physics and Mathematics Program to Drive Transformation (FoPM), a World-leading Innovative Graduate Study (WINGS) Program, the University of Tokyo, Iwadare Scholarship Foundation, LINE Yahoo Corporation, and JSPS KAKENHI Grant Number JP23KJ0726.
N.K. was supported by the Japan Society for the Promotion of Science through Grant-in-Aid for Scientific Research 21H04490.

The Hyper Suprime-Cam (HSC) collaboration includes the astronomical communities of Japan and Taiwan, and Princeton University. The HSC instrumentation and software were developed by the National Astronomical Observatory of Japan (NAOJ), the Kavli Institute for the Physics and Mathematics of the Universe (Kavli IPMU), the University of Tokyo, the High Energy Accelerator Research Organization (KEK), the Academia Sinica Institute for Astronomy and Astrophysics in Taiwan (ASIAA), and Princeton University. Funding was contributed by the FIRST program from the Japanese Cabinet Office, the Ministry of Education, Culture, Sports, Science and Technology (MEXT), the Japan Society for the Promotion of Science (JSPS), Japan Science and Technology Agency (JST), the Toray Science Foundation, NAOJ, Kavli IPMU, KEK, ASIAA, and Princeton University. 

This paper makes use of software developed for Vera C. Rubin Observatory. We thank the Rubin Observatory for making their code available as free software at \url{http://pipelines.lsst.io/}.

This paper is based on data collected at the Subaru Telescope and retrieved from the HSC data archive system, which is operated by the Subaru Telescope and Astronomy Data Center (ADC) at NAOJ. Data analysis was in part carried out with the cooperation of Center for Computational Astrophysics (CfCA), NAOJ. We are honored and grateful for the opportunity of observing the Universe from Maunakea, which has the cultural, historical and natural significance in Hawaii. 

The Pan-STARRS1 Surveys (PS1) and the PS1 public science archive have been made possible through contributions by the Institute for Astronomy, the University of Hawaii, the Pan-STARRS Project Office, the Max Planck Society and its participating institutes, the Max Planck Institute for Astronomy, Heidelberg, and the Max Planck Institute for Extraterrestrial Physics, Garching, The Johns Hopkins University, Durham University, the University of Edinburgh, the Queen’s University Belfast, the Harvard-Smithsonian Center for Astrophysics, the Las Cumbres Observatory Global Telescope Network Incorporated, the National Central University of Taiwan, the Space Telescope Science Institute, the National Aeronautics and Space Administration under grant No. NNX08AR22G issued through the Planetary Science Division of the NASA Science Mission Directorate, the National Science Foundation grant No. AST-1238877, the University of Maryland, Eotvos Lorand University (ELTE), the Los Alamos National Laboratory, and the Gordon and Betty Moore Foundation.

%

\section*{Data Availability}
The source code of PCFNet is available at the following GitHub repository: \url{https://github.com/YoshihiroTakeda/PCFNet}. The materials are provided under the MIT license.

\vspace{5mm}
\facilities{Subaru(HSC)}

\software{
astropy \citep{Astropy2013,Astropy2018},
Numpy \citep{Harris2020},
Matplotlib \citep{Hunter2007},
Pandas \citep{Mckinney2010},
Scipy \citep{Virtanen2020},
Pytorch \citep{Paszke2019},
Torchvision \citep{Torchvision2016},
Pyro \citep{Bingham2019},
Dask \citep{Dask2016},
Scikit-learn \citep{Pedregosa2011},
uncertainties \citep{Lebigot2016}
          }



\bibliography{pcf_main}{}
\bibliographystyle{aasjournal}


\end{document}